\documentclass[apj]{emulateapj}
\usepackage{times}
\usepackage[backref,breaklinks,colorlinks,citecolor=cyan]{hyperref} 
\usepackage{natbib}
\usepackage{booktabs}
\usepackage{graphicx}
\usepackage{multirow}
\usepackage{amsmath}
\usepackage[utf8x]{inputenc}

\shorttitle{DASH: The Size-Mass Relation}
\shortauthors{Mowla et al.}
\slugcomment{Submitted to the Astrophysical Journal}

\newcommand{\totgal}{910}
\newcommand{\totacs}{748}
\newcommand{\totdash}{162}

\begin{document}

\title{COSMOS-DASH: the Evolution of the Galaxy Size-Mass Relation Since $\MakeLowercase{z}\sim$ 3 from New Wide Field WFC3 Imaging Combined with CANDELS/3DHST}

\author{Lamiya A.\ Mowla\altaffilmark{1}, Pieter van Dokkum\altaffilmark{1}, Gabriel B.\ Brammer\altaffilmark{2,3}, Ivelina Momcheva\altaffilmark{2},Arjen van der Wel\altaffilmark{4,5}, Katherine Whitaker\altaffilmark{6,7}, Erica Nelson\altaffilmark{7}, Rachel Bezanson\altaffilmark{8}, Adam Muzzin\altaffilmark{9},   Marijn Franx\altaffilmark{10}, John MacKenty\altaffilmark{2}, Joel Leja\altaffilmark{7,11}, Mariska Kriek\altaffilmark{12}, Danilo Marchesini\altaffilmark{13}}
\email{lamiya.mowla@yale.edu}

\altaffiltext{1}{Astronomy Department, Yale University, New Haven, CT 06511, USA}
\altaffiltext{2}{Space Telescope Science Institute, 3700 San Martin Dr, Baltimore, MD 21211}
\altaffiltext{3}{Cosmic Dawn Center, Niels Bohr Institute, University of Copenhagen, Juliane Maries Vej 30, DK-2100 Copenhagen $\phi$, Denmark}
\altaffiltext{4}{Max-Planck-Institut f¨ur Astronomie, K¨onigstuhl 17, D-69117, Heidelberg, Germany}
\altaffiltext{5}{Sterrenkundig Observatorium, Universiteit Gent, Krijgslaan 281 S9, B-9000 Gent, Belgium}
\altaffiltext{6}{Department of Physics, University of Connecticut, Storrs, CT 06269, USA}
\altaffiltext{7}{Cosmic Dawn Center (DAWN), Niels Bohr Institute, University of Copenhagen / DTU-Space, Technical University of Denmark}
\altaffiltext{8}{Harvard-Smithsonian Center for Astrophysics, 60 Garden Street, Cambridge, MA}
\altaffiltext{9}{University of Pittsburgh, Department of Physics and Astronomy, 100 Allen Hall, 3941 O’Hara St, Pittsburgh PA 15260, USA}
\altaffiltext{10}{Department of Physics and Astronomy, York University, 4700 Keele St., Toronto, ON MJ3 1P3, Canada}
\altaffiltext{11}{Leiden Observatory, P.O. Box 9513, 2300 RA, Leiden, The Netherlands}
\altaffiltext{12}{NSF Astronomy and Astrophysics Postdoctoral Fellow}
\altaffiltext{13}{Astronomy Department, University of California, Berkeley, CA}
\altaffiltext{14}{Department of Physics and Astronomy, Tufts University, Medford, MA}

\begin{abstract}
We present COSMOS-Drift And SHift (DASH), a Hubble Space Telescope WFC3 imaging survey of the COSMOS field in the $H_{160}$ filter. The survey comprises 456 individual WFC3 pointings corresponding to an area of 0.49 deg$^2$ (0.66 deg$^2$ when including archival data) and reaches a $5\sigma$ point-source limit of $H_{160}=25.1$ (0\farcs3 aperture). COSMOS-DASH is the widest HST/WFC3 imaging survey in $H_{160}$ filter, tripling the extragalactic survey area in the near-infrared at HST resolution. We make the reduced $H_{160}$ mosaic available to the community. We use this dataset to measure the sizes of \totdash\ galaxies with $\log(M_{\star}/M_{\odot})>11.3$ at $1.5<z<3.0$, and augment this sample with \totacs\ galaxies at $0.1<z<1.5$ using archival ACS imaging. We find that the median size of galaxies in this mass range changes with redshift as $\langle{}r_{\rm eff}\rangle=(10.4\pm 0.4)\times(1+z)^{(-0.65\pm 0.05)}$\,kpc.  Separating the galaxies into star forming and quiescent galaxies using their restframe $U-V$ and $V-J$ colors, we find no statistical difference between the median sizes of the most massive star-forming and quiescent galaxies at $\langle z\rangle =2.5$: they are $4.9 \pm 0.9$ kpc and $4.3 \pm 0.3$ kpc respectively.  
However, we do find a significant difference in the S\`ersic index between the two samples, such that massive quiescent galaxies have higher central densities than star forming galaxies.
We extend the size-mass analysis to lower masses by combining it with the 3D-HST/CANDELS sample of \citet{VanderWel2014a}, and derive empirical relations between size, mass, and redshift. Fitting a relation of the form $r_{\rm eff}=A \times m_{\star}^{\alpha}$, with $m_{\star}=M_{\star}/5\times 10^{10}\,M_{\odot}$ and $r_{\rm eff}$ in kpc, we find $\log A=-0.25 \log$ $(1+z)+0.79$ and $\alpha=-0.13\log (1+z)+0.27$. We also provide relations for the subsamples of star forming and quiescent galaxies. Our results confirm previous studies that were based on smaller samples or ground-based imaging.
\end{abstract} 


\section{Introduction}
The sizes of galaxies reflect their assembly histories and their connection to their dark matter halos \citep{Mo1997TheDisks,Kravtsov2012TheGalaxies,Jiang2018IsSize}. Different modes of assembly of stars in galaxies lead to a different growth of their radii: passive evolution will cause no significant growth in size or mass, but only a maturation of the existing stellar population; dry major mergers lead to a proportional growth in size and mass as the two bodies come to dynamic equilibrium; and dry minor mergers increase the size of galaxies more rapidly by building an outer envelope \citep{Bezanson2009,Naab2009MinorGalaxies}. When gas physics are considered the evolution can be more complex; e.g., ``wet" gas-rich mergers may trigger compact starbursts leading to larger post-merger disks \citep{Hernquist1989TidalGalaxies,Robertson2005AFormation}, while gas flows to the central regions may both form compact bulges and feed a central black hole \citep{Efstathiou1982TheGalaxies,Dekel2013WetNuggets,Barro2017SpatiallyObservations}. The size of a galaxy may also hold information on the properties of the dark matter halo; galaxy size may be proportional to the halo virial radius as a result of conservation of angular momentum during the collapse and cooling of a galaxy \citep{Mo1997TheDisks,Dutton2006AExpansion,Shankar2011SizeUniverse,Kravtsov2012TheGalaxies,Porter2014UnderstandingGalaxies,Somerville2017a}, although it is unclear whether this expected correlation is actually preserved in the galaxy formation process \citep{DeFelippis2017TheSimulation, Jiang2018IsSize}.

Observationally, the sizes of galaxies have been found to vary significantly with galaxy mass, color (or star formation activity) and redshift. Generally, the sizes are larger for galaxies that are more massive, galaxies that are forming stars, and galaxies at lower redshift
\citep{Kormendy1996,Shen2003TheSurvey,Ferguson2003TheGalaxies,Trujillo2005TheFIRES,Elmegreen2007ResolvedRedshift,Williams2009TheZ=2,Mosleh2017Connection=2,Ono2012Evolution7,VanderWel2014a,Bernardi2012SystematicMorphology,Carollo2013,Lange2014GalaxyZ}. 
At intermediate masses, the slope of the size-mass relation is found to be shallow for star forming 
galaxies ($r_{\rm eff} \propto M^{0.2}_{\star}$)
and steeper for quenched galaxies ($r_{\rm eff} \propto M^{0.8}_{\star}$), where $r_{\rm eff}$ is the half-light radius. Both galaxy types exhibit a large intrinsic scatter in the size-mass relation at all redshifts \citep{VanderWel2014a}.

The slope of the size-mass relation of star forming galaxies is similar to the growth track of individual galaxies, both in observations and simulations \citep{Lilly1997HubbleZ=1,Ravindranath2004EvolutionDistribution,Trujillo2005TheFIRES,VanDokkum2015_compactmassive}. Following quenching galaxies follow a steeper growth track in the size-mass plane, probably because dry minor mergers rapidly increase the size \citep[see][]{Hilz2012HowFraction,Carollo2013,VanDokkum2015_compactmassive}. Physically,  the central stellar density has been proposed as a key parameter connecting galaxy morphology and star formation histories \citep{Bezanson2009,Carollo2013,Fang2013AGalaxies,VanDokkum2014_densecore,Whitaker2016}. Galaxies with high central densities are found to be redder with lower specific star formation rate than bluer galaxies at a given redshift. A possible explanation is that feedback mechanisms that shut off star formation are more effective when the central density becomes high \citep[e.g.,][]{Croton2005TheGalaxies,Conroy2014PreventingHeating}.

Whether the same processes operate in the most massive galaxies, here defined as galaxies with $M_{\rm\star}>2\times 10^{11}$\,M$_{\odot}$, is still an outstanding question. \citet{Carollo2013} present a comprehensive analysis of the sizes of galaxies at 0.2$<$z$<$1 in the COSMOS field, measured from the ACS F814W imaging. Out to $z\sim 1$ nearly all such galaxies are found to be quiescent \citep{Hahn2014PRIMUS:0.8}. At higher redshift this mass range is not commonly studied, as their number is low in the fields that have been observed so far with \textit{Hubble Space Telescope} (HST) in the near-IR. \citet{VanderWel2014a} studied the mass-size relation in the 3D-HST/CANDELS fields, and finds that the most massive star forming and quiescent galaxies at $z\sim 2.5$ have similar sizes. That is, the ``rule'' that star forming galaxies are larger than quiescent galaxies appears to not apply at the highest masses and redshifts. However, given the small number of galaxies with masses $>2\times 10^{11}$\,M$_{\odot}$ found within extragalctic pencil-beam studies, this is largely driven by the extrapolation of trends seen at lower masses; essentially, the fitted relations to lower mass galaxies intersect at $M_* \sim 5\times 10^{11}$\,M$_{\odot}$.  Recently \citet{Faisst2017} studied the sizes of galaxies in this mass range out to $z\sim 2$ using ground-based imaging, calibrated with HST data in smaller fields. They find similar results as \citet{VanderWel2014a}. Similarly, \citet{Hill2017} study the size evolution of galaxies since $z\sim 5$ using number density-matched samples, again consistent with previous size measurements in smaller fields.

Here we build on these previous studies by studying the most massive galaxies out to $z\sim 3$ with a new wide-field HST survey,
COSMOS-Drift And SHift (DASH). COSMOS-DASH provides the large area and high resolution needed for structural study of massive galaxies at $1.5 \leq z \leq 3.0$. It is a wide and medium depth survey using the near-infrared channel of Wide Field Camera 3 (WFC3) on HST, utilizing a novel drift-and-shift technique. COSMOS-DASH covers 0.49 deg$^2$ of the UltraVISTA \citep{McCracken2012} deep stripes in the COSMOS field down to $H_{160}=25.1$, or 0.66 deg$^2$ when archival data are included \ref{tab:archival},
tripling the extragalactic survey area observed by HST in the near-IR \citep{Momcheva2016a}. 

The paper is structured as follows. In Section \ref{sec:dash} we give a brief description of the COSMOS-DASH survey.  In Section \ref{sec:sample}, the selection of the massive galaxy sample and the separation of quenched galaxies from the star forming galaxies are described. Section \ref{sec:size} goes into the details of the size measurement of galaxies using COSMOS-DASH images. Analysis of the evolution of size-mass relation is described in Section \ref{sec:evolution}, while in Section \ref{sec:discussion} we interpret the results in the context of the termination of star formation in the most massive galaxies.

In this paper, we assume a $\Lambda$CDM cosmology with $\Omega_{\rm m}=$ 0.3, $\Omega_{\Lambda}=$ 0.7, and $H_0=$ 70 km s$^{-1}$ Mpc$^{-1}$. 

\section{COSMOS-DASH}
\label{sec:dash}
\begin{figure*}[ht]
\centering
\includegraphics[width=\textwidth]{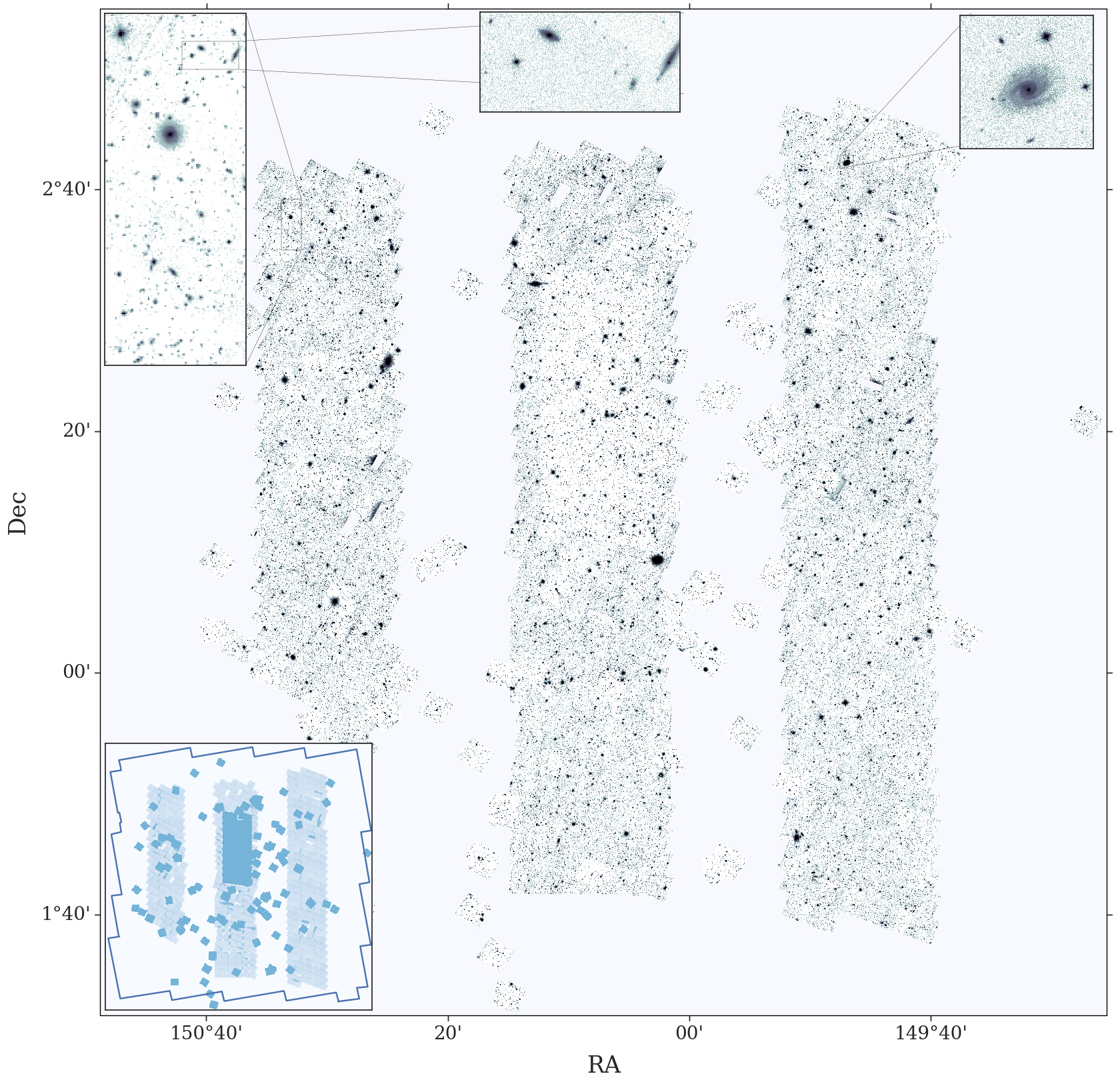}
\caption{The COSMOS-DASH $H_{160}$ mosaic. The science image is shown in the main figure, along with the exposure map in the colored inset. Zoomed-in portions of the science image are shown as well. The area contributed by COSMOS-DASH is 0.49\,deg$^2$; the total area with $H_{160}$ data (including the deeper CANDELS imaging and various other archival data sets listed in Table \ref{tab:archival}) is 0.66\,deg$^2$.  The magenta outline shows the 1.64 degree$^2$ covered by ACS $I_{814}$ \citep{Koekemoer2007TheProcessing}.}
\label{fig:dash_wht}
\end{figure*}

Wide-field near-infrared (IR) surveys have proven invaluable for the study of the high mass end of the galaxy mass function at $z>1$ (where the rest-frame optical emission shifts into the near-IR) and for determining the prevalence of short-lived events such as mergers, the properties and demographics of AGN, and the evolution of galaxy groups and clusters. Such surveys have been undertaken from the ground (e.g. NMBS \citep{Whitaker_2011}, UltraVISTA \citep{McCracken2012,Muzzin2013THESURVEY}, UKIDSS-UDS \citep{Lawrence2007, Williams2009TheZ=2}) but so far not with the \textit{Hubble Space Telescope} (HST). The largest area imaged with HST in the optical ($I_{814}$) is the 1.7 deg$^2$ COSMOS field in 640 orbits 
\citep{Scoville2006TheOverview}, whereas the largest area imaged in the near-IR ($J_{125}$ and $H_{160}$) is the 0.25 deg$^2$ CANDELS survey \citep[900 orbits;][]{Koekemoer2007TheProcessing}.
Until now HST near-IR surveys over larger areas have not been done because it is very inefficient to observe multiple pointings within a single HST orbit. We have developed a technique to circumvent this limitation, enabling an order of magnitude increase in the efficiency of large area mapping with HST.

The Drift And SHift technique greatly increases the efficiency of mapping large areas as it requires only a single guide star acquisition per pointing. In the COSMOS-DASH survey, we imaged 0.49 deg$^2$ of the COSMOS field in H$_{160}$ in 57 orbits (31 arcmin$^2/$orbit) reaching a depth of $H_{160}=25.1$ (0\farcs3 aperture).

\subsection{Drift And SHift (DASH)}

\begin{figure*}[t]
\centering
\includegraphics[width=\textwidth]{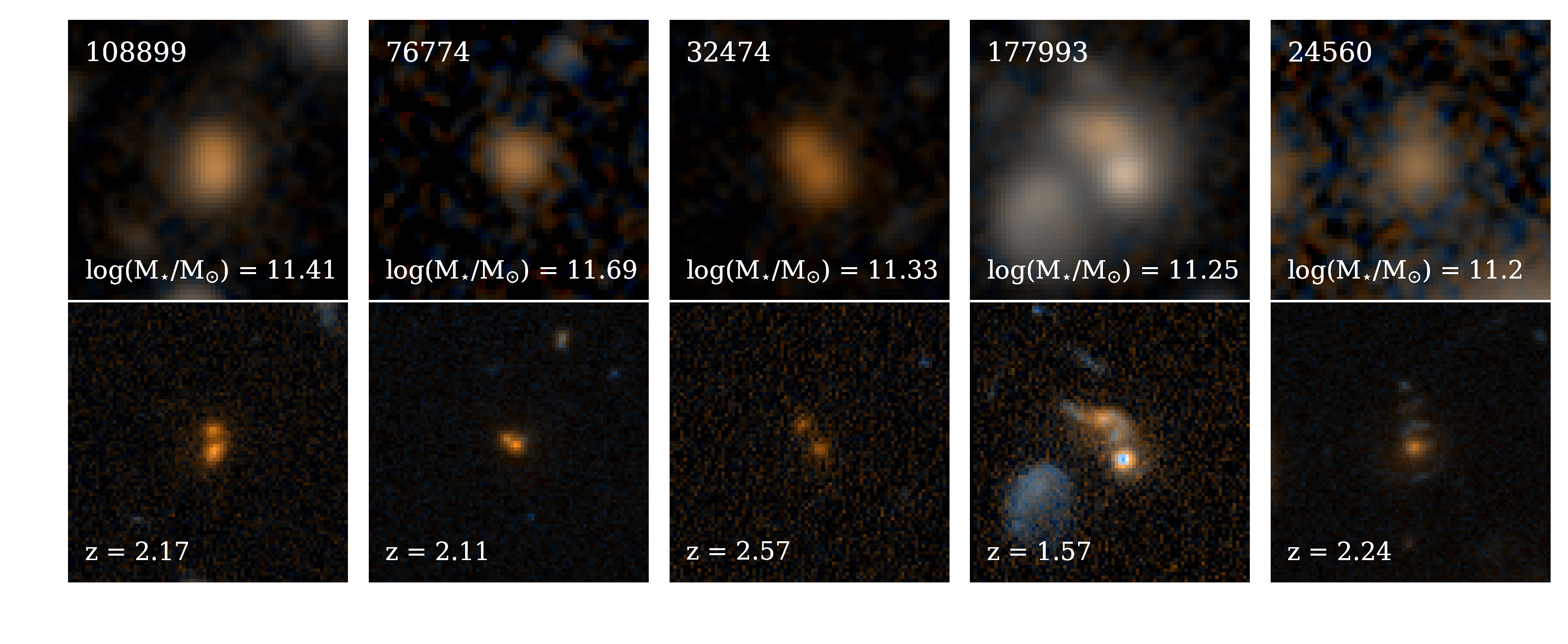}
\caption{Images of massive galaxies at z$>$1.5 observed with UltraVISTA (top panel) and COSMOS-DASH (bottom panel). The UltraVISTA images are created using the UVISTA $H$ and $i$ band images. The COSMOS-DASH images are created using the $H_{160}$ and $I_{814}$ images. Each panel is $8\arcsec\times8\arcsec$. The first three galaxies are close pairs in the COSMOS-DASH image which appear as single object in the UltraVISTA image.} 
\label{fig:dash_uvista_compare}
\end{figure*}

Drift And SHift (DASH) is a new technique for efficient large area observations using the near-infrared (IR) channel of Wide Field Camera 3 (WFC3) on the \textit{Hubble Space Telescope} \citep{Momcheva2016a}. In standard HST observations guide stars are acquired for each new pointing. Acquiring a guide star takes approximately 10 minutes, which means that short exposures are only possible with very large overheads and the total number of pointings that can be obtained within an orbit's visibility window is small\footnote{HST is limited to two acquisitions per orbit by policy.}. 
This limitation can be circumvented by acquiring guide star for only the first pointing and guiding with the three HST gyros for the rest of the pointings. This makes it possible to observe up to 8 WFC3 pointings in a single orbit, greatly increasing HST's large scale mapping capabilities.

During a standard guided exposure, the three HST gyros receive continuous corrections from the Fine Guidance Sensors (FGS). Turning off guiding stops the stream of corrections from the FGS and the telescope begins to drift with an expected rate of $0\farcs 001 - 0\farcs 002$ per second. 
In CCDs this would lead to detrimental smearing of the image in a typical 5 minute exposure. However,
the WFC3/IR detector can perform multiple non-destructive, zero overhead reads throughout the exposure. By setting the time  between reads to 25 seconds or less, the drift in between reads is $\leq 0\farcs 05$ - less than half a pixel (pixel scale $=0\farcs129$).   The  data obtained in between the reads can be treated as independent 25\,s exposures, that can be drizzled to restore the full resolution of WFC3. In \citet{Momcheva2016a} we demonstrated that the resolution of the WFC3 camera is preserved in this process, and that structural parameters of the galaxies are consistent with those measured in guided observations.

\subsection{Observations and Data Reduction}

The ``Drift And SHift" (DASH) method was used in the Cycle 23 COSMOS-DASH program (Program ID: GO-14114) to obtain 456 WFC3 $H_{160}$ pointings in 57 orbits, covering an area of 0.49 degree$^2$ in the COSMOS field.
The data were obtained between November 2016 and June 2017. The data were reduced by constructing 11 or 12 individual 25\,s exposures
from the differences between subsequent reads within each science exposure. The full set of $\sim 5000$ exposures, together with all other existing $H_{160}$ data in the COSMOS field, were then drizzled into a single, large mosaic.
The details of the reduction procedure are described in \citet{Momcheva2016a} and in Appendix \ref{ap:reduction}. 

The final mosaic of the COSMOS-DASH image is $50,000\times 50,000$ pixels (with $0\farcs 1$\,pix$^{-1}$), centered at RA$=$10:00:28.6, DEC$=+$02:12:21.0. It is shown in Fig.\ 1 and is available from the COSMOS-DASH website\footnote{\url{http://www.stsci.edu/~imomcheva/data/COSMOS_DASH/}}.
The total area of the $H_{160}$ imaging is 0.66 deg$^2$, with COSMOS-DASH contributing 0.49 deg$^2$ and the remainder archival data taken in a variety of programs (list given in the appendix). This roughly triples the area of extragalactic blank survey fields that have been imaged in the near-IR with HST.  We note that of these additional programs, our mosaic subsumes the data from GO-12990 (PI: Muzzin) which obtained targeted imaging of 12 galaxies with $\log M_*>11.6$ at 1.5 $<$ z $<$ 3.0 in COSMOS with $\sim$1000 second integration time.

\section{Sample of Massive galaxies}
\label{sec:sample}

\subsection{UVISTA Catalog}
\label{sec:uvista}

We use the UitraVISTA catalog of \citet{Muzzin2013THESURVEY} for our sample selection in the COSMOS field. 
Objects are selected from UltraVISTA $K_s$ band imaging that reaches a depth of $K_{\rm s,tot}=23.4$ AB at 90\,\% completeness. The catalog contains PSF-matched photometry in 30 photometric bands covering the wavelength range 0.1$\mu$m $\to$ 24$\mu$m and includes the available GALEX, CFHT/Subaru, UltraVISTA, and S-COSMOS dataset. Each galaxy in the catalog has a photometric
redshift determined by fitting the photometry in the 0.1$\mu$m $\to$ 8.0$\mu$m bands to template Spectral Energy Distribution (SEDs) using the EAZY code \citep{Brammer2008}. EAZY also provides rest-frame U, V and J colors which we used to separate the star-forming and quiescent galaxies. We note that \citet{Muzzin2013THESURVEY} used the default EAZY template set, and did not include the ``old and dusty" template used in \citet{Skelton20143D-HSTMasses}. This template was designed to extend the red boundary in the UVJ color-color space (see \ref{fig:uvj}). Without spectroscopic redshifts for the reddest galaxies at the highest redshifts it is difficult to determine which template set provides the most accurate description of the galaxy population.
Stellar masses for all galaxies in the catalog were determined by fitting the SEDs of galaxies to stellar population synthesis (SPS) models using the FAST code \citep{Kriek2009AnZ=2.2} using \citet{Bruzual2003Stellar2003} templates with solar metallicity, a wide range in age, an exponentially declining star formation history, a \citet{Chabrier2003} IMF and a \citet{Calzetti1999TheGalaxies} dust extinction function. Details of the catalog can be found in \citet{Muzzin2013THESURVEY}.

Although this catalog is based on an early release of the near-IR data in this field (UltraVISTA-DR1) it is easily deep enough for the relatively bright galaxies analyzed in this paper. In Appendix \ref{ap:flux_correct} we describe two corrections that are applied to the masses. We first correct them for flux that is missing in the catalog aperture, using the GALFIT total magnitudes. Next we determine whether any systematic offsets need to be applied. We show that the stellar masses (and photometric redshifts) are in very good agreement with the recent COSMOS2015 catalog (based on DR2; \citep{Laigle2016}). However, the masses are 0.1 dex {\em lower} than those in \citet{VanderWel2014a}, for the same galaxies. The same offset is obtained when matching the number density of galaxies with masses $M_*>10^{11}$\,M$_{\odot}$ in the two catalogs. As discussed in the Appendix we apply a 0.1 dex offset to all the masses, for consistency with \citet{VanderWel2014a}.

\subsection{Selection of Galaxies for Size Analysis}

\begin{figure}[t]
\centering
\includegraphics[width=0.46\textwidth]{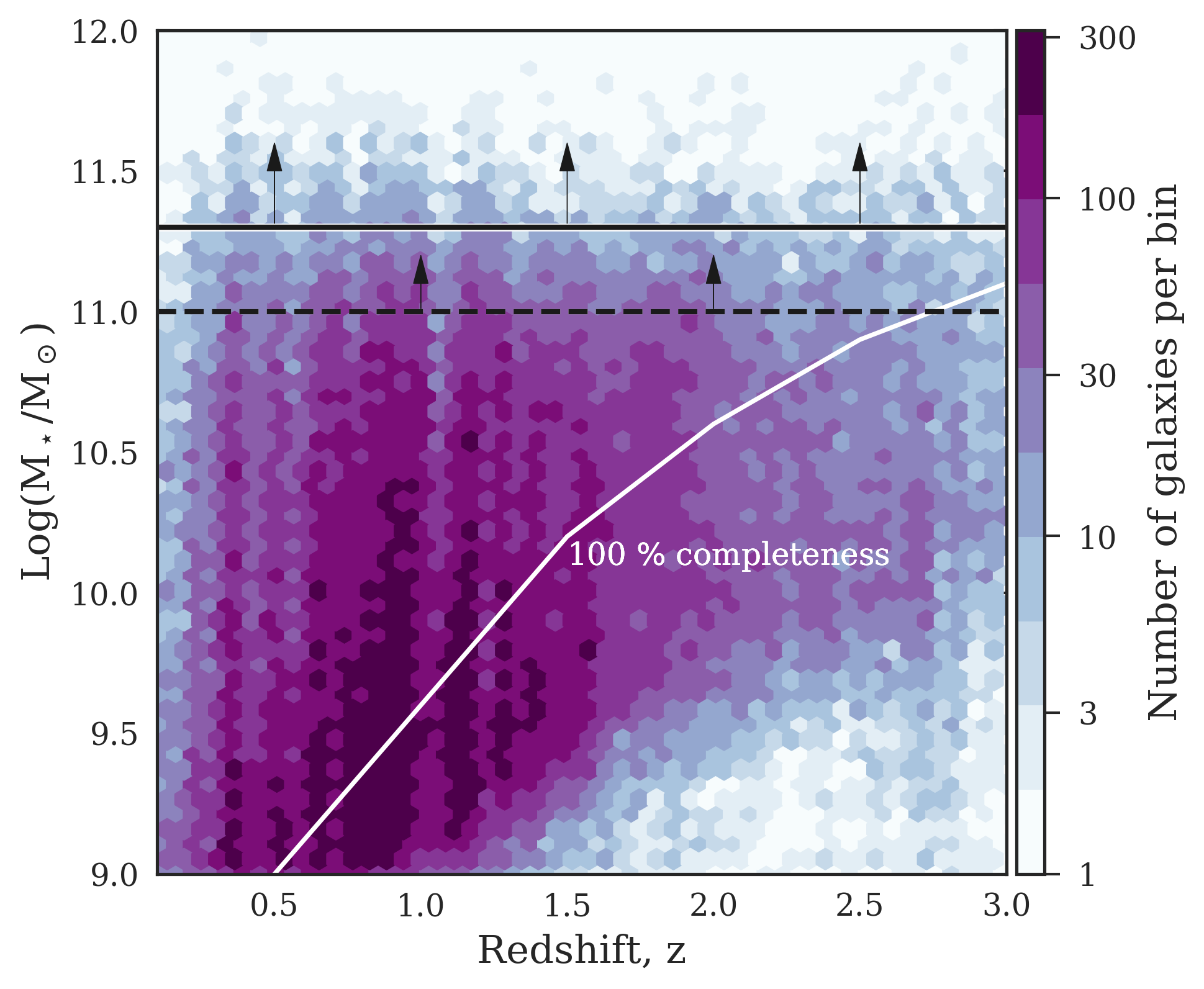}
\caption{Sample selection from the UltraVISTA catalog \citet{Muzzin2013THESURVEY}. The entire UltraVISTA sample with $H<21.5$ is shown.  The white line shows the 100\,\% mass-completeness limit from \citet{Muzzin2013THESURVEY}.
}
\label{fig:mass_cut}
\end{figure}

In this paper we study the most massive galaxies at $0.1\leq z_{\rm phot} \leq 3.0$ with $M_{\star}\geq 2 \times10^{11} M_{\odot}$. 
The existing 3D-HST/CANDELS samples are sufficiently large for determining the size-mass relation below this limit \citep{VanderWel2014a}; furthermore, at lower masses the COSMOS-DASH imaging is too shallow for accurate measurements of structural parameters at the highest redshifts. Figure \ref{fig:mass_cut} shows all galaxies in UltraVISTA in the plane of mass versus redshift, before the corrections are applied. Our sample is well above the completeness limits of the catalog, even at $z=3$. We find \totgal\ galaxies in the catalog with $M_{\rm \star,corr}>2\times10^{11}M_{\odot}$, of which \totacs\ galaxies are at $0.1<z<1.5$ and \totdash\ galaxies are at $1.5<z<3.0$.

\begin{figure}[t]
\centering
\includegraphics[width=0.46\textwidth]{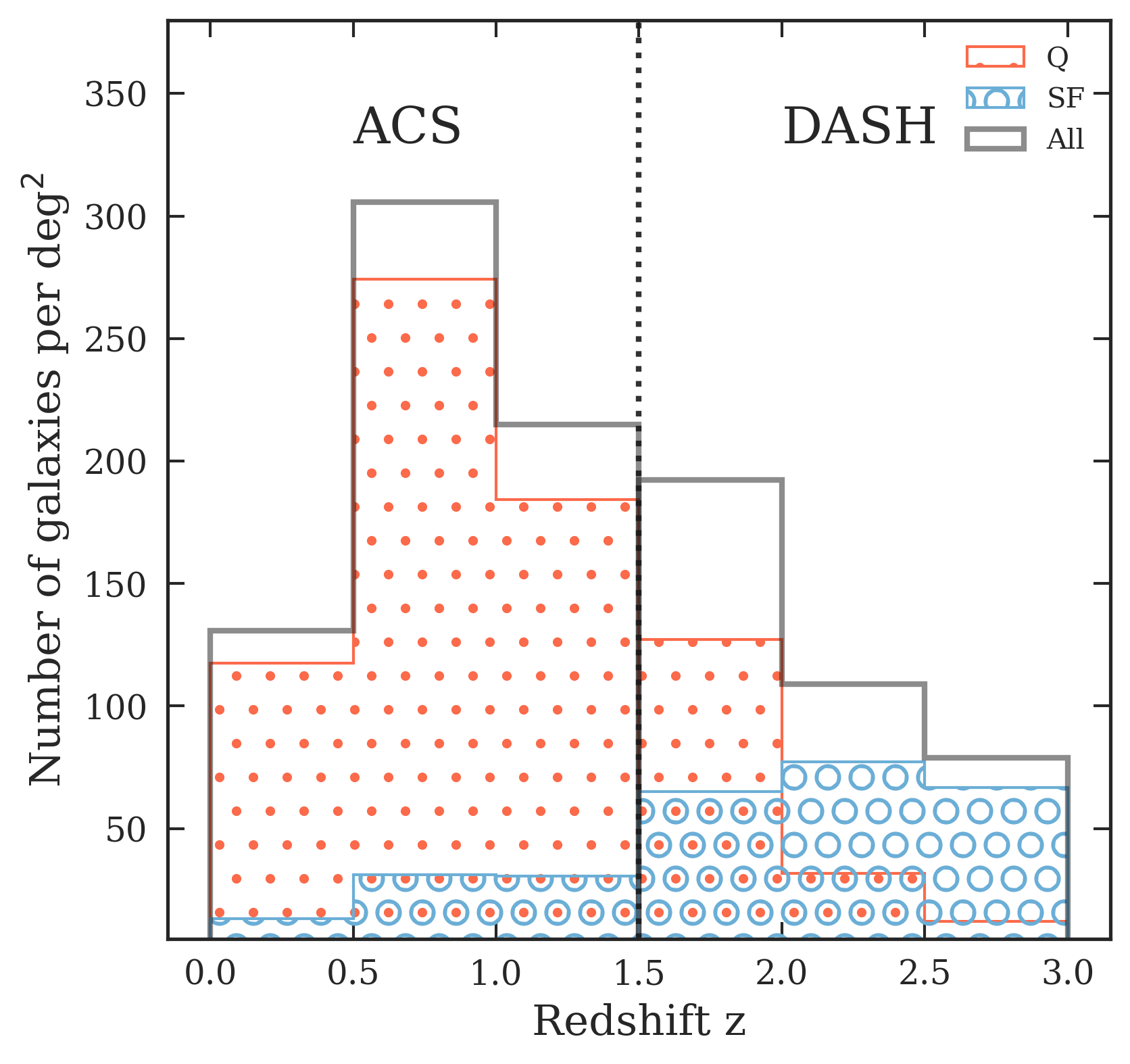}
\caption{The number of massive galaxies
with $\log M_*>11.3$ at $0.1<z<3.0$ per deg$^2$ for which a stable fit has been found. The star-forming and quiescent galaxies are shown in blue and red respectively, while the entire sample is shown in grey. Sizes of galaxies at $z>1.5$ are measured from the COSMOS-DASH image ($H_{160}$) and from ACS  ($I_{814}$) imaging at  $z<1.5$.}
\label{fig:mass_hist}
\end{figure}

The 0.66 deg$^2$ COSMOS-DASH mosaic does not cover the entire UltraVISTA area. Furthermore, at $z\sim 2$
the $H_{160}$ filter samples the rest-frame $V$ band, but at $z\sim 1$ it samples the rest-frame $I$ band. Because of these considerations we use the COSMOS-DASH data only in the redshift range
$1.5<z<3.0$. For galaxies at $0.1<z<1.5$ we use the HST ACS $I_{814}$ observations of the COSMOS field \citep{Koekemoer2007TheProcessing}. The ACS mosaic spans $1.66$\,deg$^2$, and covers the entire UltraVISTA area. By combining the ACS and WFC3 imaging we ensure that the volumes at $z\sim 1$ and $z\sim 2$ are roughly matched, and that we measure sizes at approximately the same rest-frame wavelengths. As discussed in appendix \ref{ap:wave_correct}, we correct for residual wavelength-dependent effects using previously-measured color gradients of galaxies.

We obtained the reduced ACS v2.0 mosaic from the NASA/IPAC Infrared Science Research Archive\footnote{\url{http://irsa.ipac.caltech.edu/data/COSMOS/images/acs_mosaic_2.0/}}. 
The image mosaic software Montage v5.0\footnote{\url{http://montage.ipac.caltech.edu/}} was used to create square cutouts of $18\arcsec \times 18\arcsec$ centered on the galaxies. We visually inspected all  galaxies which are covered by the images and remove those that are on an edge. A total of 203 galaxies at $1.5<z<3.0$ are covered by the 0.66 deg$^2$ COSMOS-DASH mosaic (308 galaxies/deg$^2$), while 788 galaxies at $0.1<z<1.5$ are covered by the 1.64 deg$^2$ ACS-COSMOS (493 galaxies/deg$^2$). These 991 galaxies at $0.1<z<3.0$ form our sample for the structural study. 

\subsection{Classifying Star Forming and Quiescent Galaxies}

As part of the analysis we split the sample into  star-forming and quiescent sub-population of galaxies. Observationally, for intermediate to massive galaxies, the sizes at a fixed stellar mass and redshift are dependent on their colors, with the bluer galaxies being larger than redder galaxies. We want to test whether this relation holds at the highest masses and redshifts. 

We use the rest-frame $U-V$ and $V-J$ color space to separate galaxies into star-forming (SF) and quiescent (Q) candidates. Galaxies occupy distinct regions in the $U-V$ vs. $V-J$ plane depending on their specific star formation rate and dust content, as demonstrated by \citet{Labbe2002, Whitaker2012}. Young galaxies with high star formation rates which are red due to high dust content occupy a different region in the $UVJ$ diagram than old quenched galaxies. In this paper we use a redshift-dependent separation line in the $UVJ$ diagram to identify ourSF and Q galaxy candidates \citep{Muzzin2013THESURVEY}. Quiescent galaxies are defined as:

\begin{align}
    U-V \textnormal{ } &> \textnormal{ } 1.3 \textnormal{ \hspace{0.2cm} and \hspace{0.2cm}  } V-J < 1.5 & \textnormal{ [all z]} \nonumber \\ 
    U-V \textnormal{ } &> \textnormal{ } 0.88\textnormal{ }(V-J) \textnormal{ }+ \textnormal{ }0.69 & [0.0<z<1.0] \label{eq:sfq} \\ 
    U-V \textnormal{ } &> \textnormal{ } 0.88\textnormal{ }(V-J) \textnormal{ }+ \textnormal{ }0.59 &  [1.0<z<3.0] \nonumber   
\end{align}

\begin{figure*}[ht]
	\centering
	\includegraphics[width=\textwidth]{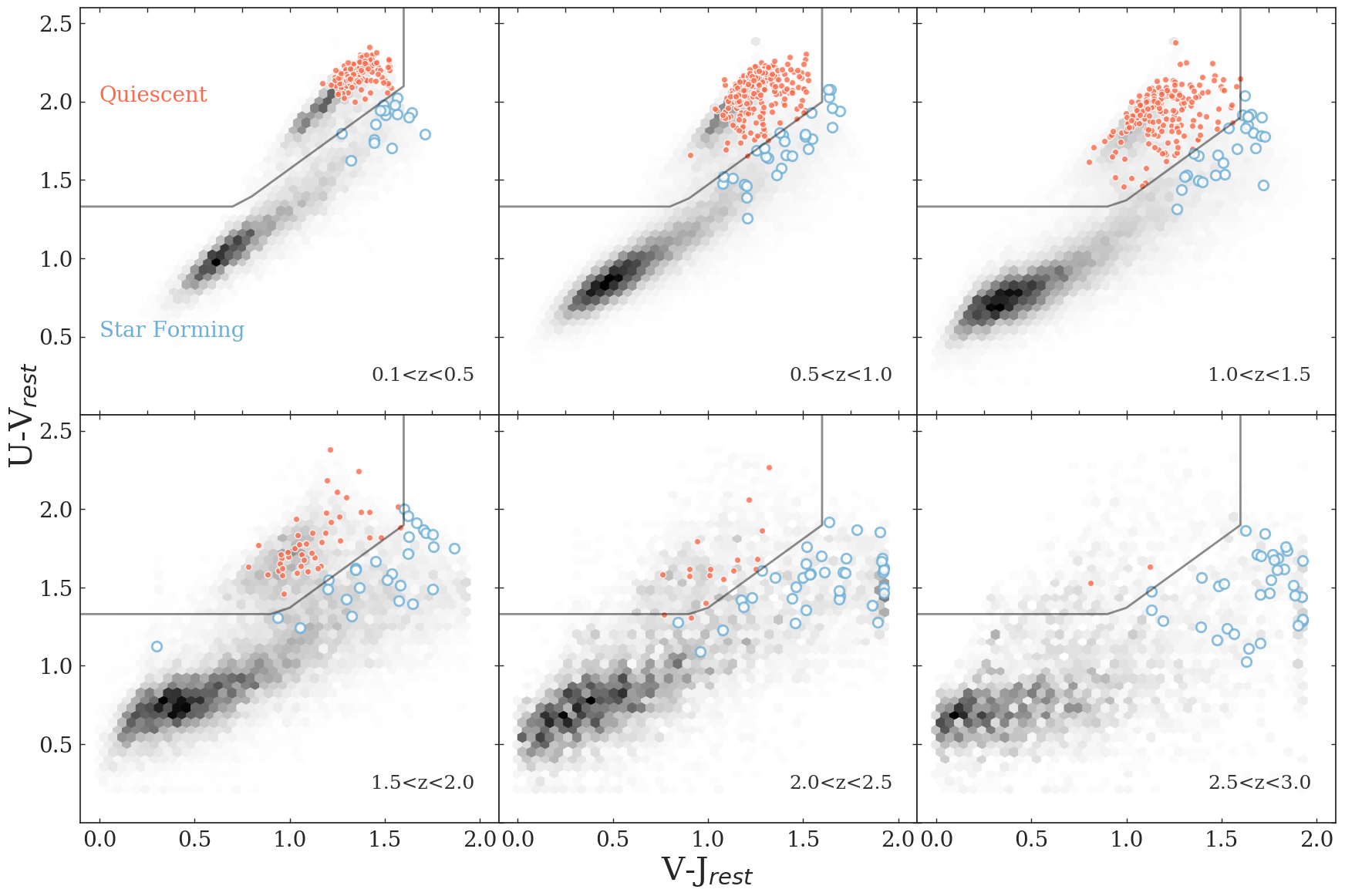}
	\caption{Rest-frame $U-V$ vs.\ $V-J$ color distribution for six redshift bins. Quiescent and star-forming galaxies are separated by the selection criteria defined in Eq. \ref{eq:sfq}, shown by the black lines. All the galaxies in the UltraVISTA catalog with  $\log(M_{\star}/{\rm M}_{\odot})>9.0$ at $0.1<z<3.0$ are shown in grey. Galaxies in our sample with $M_*>2\times10^{11}$\,M$_{\odot}$ are shown in colors: star-forming galaxies are shown as blue rings and quiescent galaxies as red circles.} 
	\label{fig:uvj}
\end{figure*}

This separation was originally defined by \citet{Williams2008DETECTIONFROM} to maximize the difference in specific star formation rates (sSFRs) between the two distinct sub-populations on the $UVJ$ diagram. The separation was adjusted by \citet{Muzzin2013THESURVEY} to fit the UVISTA sample since the rest-frame color distribution is different from the \citet{Williams2008DETECTIONFROM} sample, presumably due to small differences in photometric methods. The distribution of our sample of galaxies in the $U-V$ vs.\ $V-J$ plane is shown in Fig. \ref{fig:uvj}. At low redshift, the galaxy samples separate into distinct sub-populations in the $UVJ$ parameter space. However, at high redshift the galaxies tend to move closer to the separation line. This may indicate that there are more transition galaxies at $z\sim 2$ than at other epochs.

There are more galaxies at  $0.1<z<1.5$ than at $1.5<z<3.0$, both because the area of ACS-COSMOS is 2.5 larger than the area of the COSMOS-DASH mosaic and because the density of massive galaxies per square degree in the low redshift bin is more than double that in the higher redshift bin. This is demonstrated in Figure \ref{fig:mass_hist}, which shows the number of galaxies per square degree in six redshift bins. The higher number of more massive galaxies at low redshift is expected from the evolution of stellar mass function of galaxies.

Figure \ref{fig:mass_hist} also shows the number of SF and Q galaxies per degree $^{2}$ at 0.1$<$z$<$3.0. At z$<$0.5 less than 10$\%$ of the total galaxies are SF,  whereas at 2.5$<$z$<$3.0 more than 85$\%$ of the galaxies are SF (see also \citet{Marchesini2014}). We will return to this in Section \ref{sec:transition}. In all the plots in this paper, ``all" galaxies are represented by grey squares and solid lines, ``quiescent" galaxies are presented by red dots and dashed lines and ``star-forming" galaxies are represented by blue circles and dotted lines, unless stated otherwise. 

\section{Size Determinations}
\label{sec:size}


We use GALFIT \citep{Peng2010} to fit one-component S\'ersic profiles to the galaxies. The effective radius ($r_{\rm eff}$)  is the semi-major axis of the ellipse containing half of the total flux of the best-fitting model. For galaxies at $0.1<z<1.5$ the $I_{814}$ ACS-COSMOS image is used and for galaxies at $1.5<z<3.0$ the $H_{160}$ COSMOS-DASH image is used.  Unless otherwise stated, both the $I_{814}$ and $H_{160}$ images are processed in the same way prior to running GALFIT on them.

\subsection{Preparation of Images}

\begin{figure*}[ht]
\centering
\includegraphics[width=\textwidth]{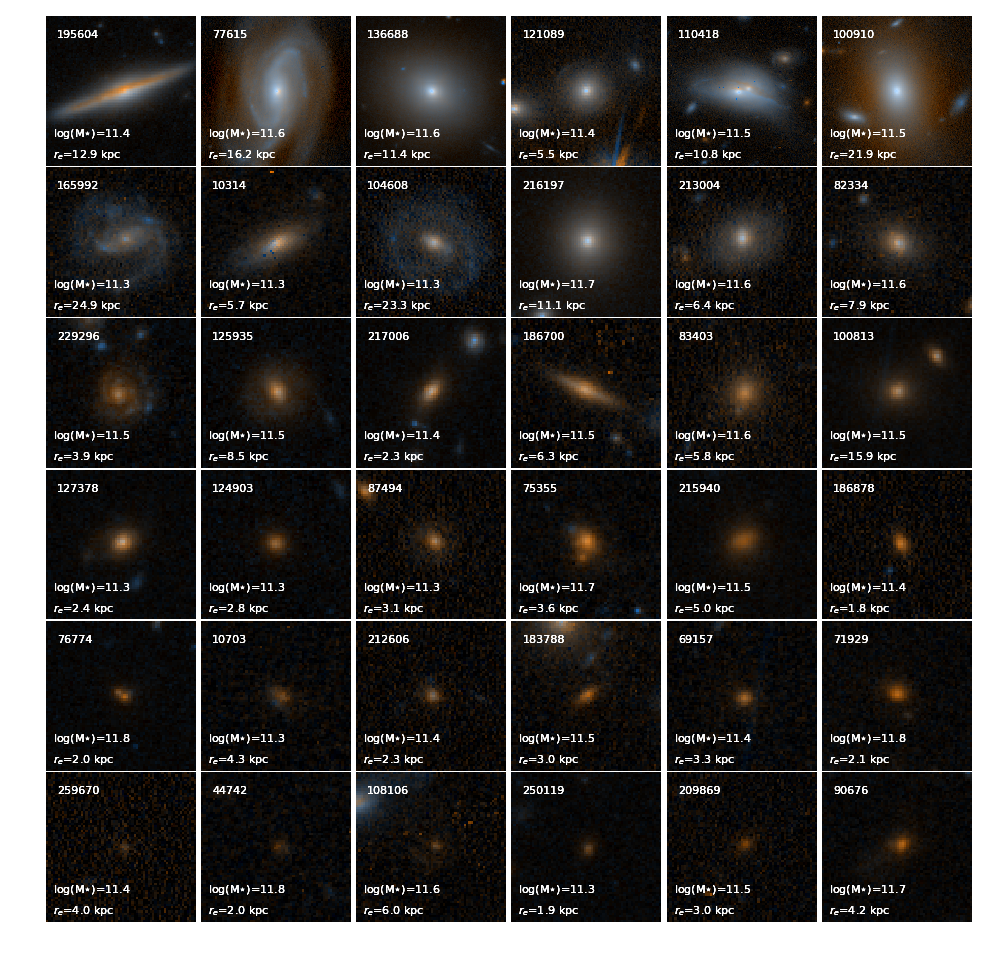}
\caption{Images of example massive galaxies created from COSMOS-DASH $H_{160}$ and ACS $I_{814}$ images. Each image is 50 kpc$\times$50 kpc. The  $\log$(M$_{\star}$/M$_{\odot}$) and photometric redshift $z$ of each galaxy from the UltraVISTA catalog are listed in the images. Redshift increases from top to bottom, and star formation rate increases from left to right within each row (that is, they are ordered by their distance from the UVJ separation line). The spectral energy distributions (SEDs) of these galaxies are shown in Fig. \ref{fig:gal_sed}.}
\label{fig:gal_image}
\end{figure*}

\begin{figure*}[ht]
\centering
\includegraphics[width=\textwidth]{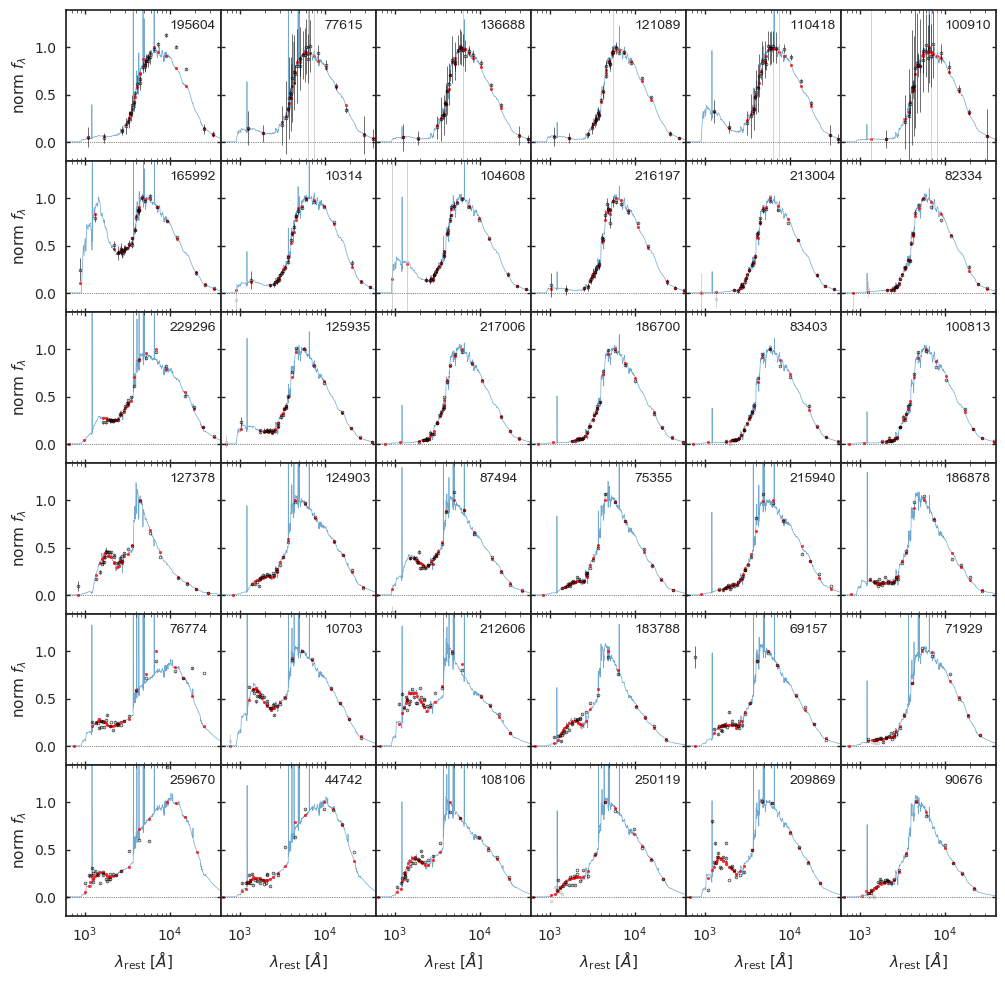}
\caption{Restframe UV to near-IR spectral energy distributions of galaxies shown in Fig. \ref{fig:gal_image}. The blue spectra are the best-fitting EAZY \citep{Brammer2008} model, the red circles show the model fluxes in the observed filters and the open black circles show the observed fluxes. } 
\label{fig:gal_sed}
\end{figure*}

The galaxy images are prepared for fitting in the following way. First, image cutouts are created from the mosaic that is appropriate for its redshift (ACS for $z<1.5$; WFC3 for $z>1.5$).  The size of the square cutout is determined in two stages. In stage 1, for each individual galaxy the image mosaic software Montage v5.0 is used to create square cutouts of sides $18\arcsec$, centered on the galaxy. We use SExtractor \citep{Bertin1996} v2.19.5 to identify all sources in the cutouts. SExtractor is run on the individual cutouts with a fixed detection threshold of 1.5 times the standard deviation above the background RMS level, 32 deblending sub-thresholds, and a minimum contrast parameter of 0.005. The size of the central object of interest is determined from the Kron radius measured by SExtractor. In stage 2, square cutouts are created with sides of length equal to 7 times the semi-major axis Kron radius (7$\times$\,A\_IMAGE\,$\times$\,KRON\_RADIUS, as expressed in SExtractor outputs).

SExtractor is run again on the individual final cutouts with the same setup as described above. We use the SExtractor segmentation map to create a mask that contains all detected objects except the galaxy of interest. Noise maps are made for individual galaxies with the same size as the final cutouts, assuming that the sky background is the dominant noise component.

\subsection{PSF Model}

Detailed knowledge of the image PSF is required for robust modeling of the galaxy morphologies with GALFIT.  Many previous studies derive an empirical PSF for a given image mosaic taken from an average of observed images of isolated stars (e.g., \citet{VanderWel2014a} ).  The COSMOS-DASH mosaic used here is generated from both the shallow DASH exposures themselves and from deeper archival observations taken under the more standard ``guided and dithered'' observation strategy.  To account for the heterogeneous depth and dither/guiding properties of the COSMOS-DASH mosaic, we adopt a new method
 for generating PSF models by using the WFC3/IR empirical PSF library from \citet{Anderson2016}.  The WFC3/IR empirical PSFs are provided for all broad band filters and with the spatial variation across the detector sampled on a 3$\times$3 grid.  Four sub-pixel center positions are provided at each of these grid points.  For each galaxy we wish to model, we first insert the appropriate empirical PSF model at the exact object location in the detector frame of each individual exposure in which that object is found.  Then we drizzle these models to the same pixel grid as the final mosaic using identical parameters (i.e., the relative image weights and pixel scale parameters).  In this way, the final model PSFs of each object\footnote{Drizzled PSFs used with GALFIT for each of our massive galaxies are made available at \url{http://www.stsci.edu/~imomcheva/data/COSMOS_DASH/}. The software used for creating the drizzled empirical PSFs is available at \url{https://github.com/gbrammer/grizli/}.} fully account for the effects of instrument orientation, pixel resampling and image weighting that together determine the PSF in the science mosaics.

\subsection{Fitting}

\begin{figure*}[ht]
\centering
\includegraphics[width=\textwidth]{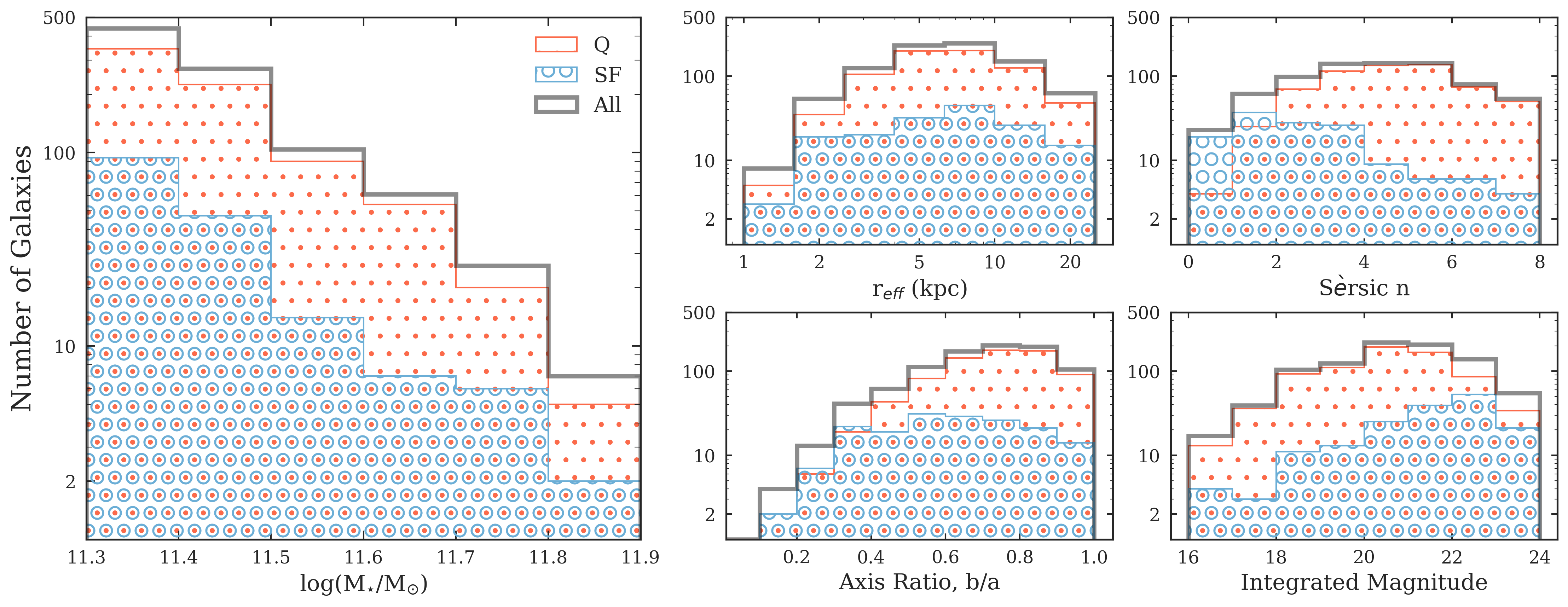}
\caption{Histograms showing the distribution of stellar mass and structural parameters of our sample galaxies. The star-forming galaxies are represented by the blue-circle and the quiescent galaxies by the red-dot hatched histograms. The grey envelopes represent the distribution of all galaxies. The left panel shows the distribution of stellar mass, the upper right panels show distributions of effective radius and S\`ersic index, and the bottom right panels show the distributions of axis ratio and of integrated magnitude.} 
\label{fig:structural_hist}
\end{figure*}

The image cutouts, along with the noise map, the appropriate PSF model, and mask, are provided to GALFIT, which is used to find the best-fitting S\'ersic model for each object. The fit parameters are total magnitude ($M$), half-light radius ($r_{\rm eff}$) measured along the semi-major axis, S\`ersic index ($n$), axis ratio ($b/a$), position angle (PA), central position ($x_0$, $y_0$) and an additive constant ($\textit{sky}$). Initial guesses for these parameters are taken from the SExtractor detection catalog that was used to create the masks. A constraints file is constructed so that GALFIT is forced to keep the S\'ersic index between 0.2 and 8, the effective radius between 0.03$\farcs$ and 40$\arcsec$ (0.3 and 400 pixels for COSMOS-DASH, 0.1$\farcs$ and 1200 pixels for ACS-COSMOS), the axis ratio between 0.0001 and 1, the magnitude between  $-3$ and $+3$ magnitudes from the input value (the SExtractor magnitude). We use a wrapper to create the GALFIT feedfiles and to run GALFIT on individual galaxy stamps. Neighboring objects in each image cutout are fit simultaneously or masked out, depending on their proximity and brightness compared to the main object: galaxies are fit simultaneously if they are less than 4 magnitudes fainter and if they are within 1$\arcsec$ from the main target (see Section \ref{sec:pair} for fitting close pairs of objects).

\subsection{Close Pairs}
\label{sec:pair}

A number of galaxies which are cataloged as a single object in the UVISTA catalog are split into pairs in the COSMOS-DASH image. Out of the 203 galaxies in our sample at $z>1.5$, 18 split into pairs. For each of the pairs, we fit the two objects simultaneously with GALFIT using the same constraints as those described above. We then estimate the mass of each of the components of the pair by dividing the total mass into two parts weighed by the flux of the components in the $H_{160}$ filter: 

\begin{equation}
    M_{\star,i} = M_{\star,\rm tot} \times \frac{F_i}{F_i+F_j}
\label{eq:pair}
\end{equation}

where $M_{\star,\rm tot}$ is the total stellar mass of the galaxy given in the UVISTA catalog and $F_i$ and $F_j$ are the total model fluxes of the two components from GALFIT. If  $M_{\star,i}>2 \times 10^{11} M_{\odot}$ then the object is kept in the sample (with the same UVISTA ID number) or else it is rejected from the sample. However, there are no pairs where both the components are above the mass cut. 14 out of the 18 have one component above the mass cut and remain in the sample with their revised mass.

\subsection{Visual Inspection and Additional Steps}
After each individual object is fit by the GALFIT wrapper, the $\chi^2$ value of the fit and the effective radius of the model are checked to ensure that the fits are reasonable. Any fits with $\chi^2 > 1.2$ or $r_{\rm eff} >$ 40 kpc are rejected and refit by tweaking the initial conditions.  Most of these ``failed" fits occur for objects with nearby bright stars, overlapping background/foreground objects or objects which are close to the edge of the mosaic with low signal to noise. In those cases the initial GALFIT estimates (based on the SExtractor analysis) may be (too) far from the true values. If a reasonable fit is obtained by changing the initial conditions or by altering the mask, the galaxy is kept in the sample. Otherwise the galaxy is rejected from our sample. 

In the second stage, all objects with $n=0.2$ or $n=8$, i.e. the boundary conditions given on the GALFIT constraint file, are visually inspected. Most of the $n=8$ objects are found to have a bright central pixel, in some cases probably due to the presence of an active nucleus. We first refit the objects following the procedure described above. If this does not resolve the problem, the galaxy is refit by fixing the S\`ersic index to the S\`ersic index determined by SExtractor. Even though SExtractor does not incorporate the PSF in its fit and the S\`ersic indices {\em should} not be reliable, in practice they correlate quite well with the GALFIT-determined ones with no obvious systematic offset. 23$\%$  of the galaxies at $0.1<z<1.5$ and 17\,\% of the  galaxies at $1.5<z<1.5$ are refit with fixed $n$.

In the third stage, each of the individual galaxies, their S\`ersic models and the subtracted background images produced by GALFIT are visually inspected. Any ``obviously'' unreasonable fit is manually refit as above. We rejected 14 galaxies because they are almost invisible in the DASH image. A total of 935 galaxies passed this final quality control step and formed the sample for our size-mass analysis. In the remainder of the paper we are describing the results for these galaxies only. The fraction of rejected objects is similar to that in previous studies. 

\subsection{Stellar Mass and Size Corrections and Final Sample}

As explained above and in the Appendix, the stellar masses are corrected so they are self-consistent with the best-fit S\`ersic model and consistent with the mass estimates from \citet{VanderWel2014a} sample. The sizes are corrected for color gradients, following the procedure of \citet{VanderWel2014a}; details are given in Appendix \ref{ap:wave_correct}. The final sample is selected such that $M_{\rm \star, corr}\geq 2 \times10^{11} M_{\odot}$. We find a total of \totgal\ galaxies at $0.1<z<3.0$ of which \totacs\ are measured from the ACS-COSMOS image ($z<1.5$) and \totdash\ are measured from the COSMOS-DASH image ($z>1.5$).

We also include the galaxies with $M_{\rm \star, vdW}\geq 2 \times10^{11} M_{\odot}$ from the CANDELS field measured by \citet{VanderWel2014a} that are not in our study. Hence the total number of massive galaxies used in the size-mass analysis are 1090, where 813 are quiescent and 277 are star-forming.

\subsection{GALFIT Error Estimation}

\begin{figure*}[ht]
\centering
\includegraphics[width=\textwidth]{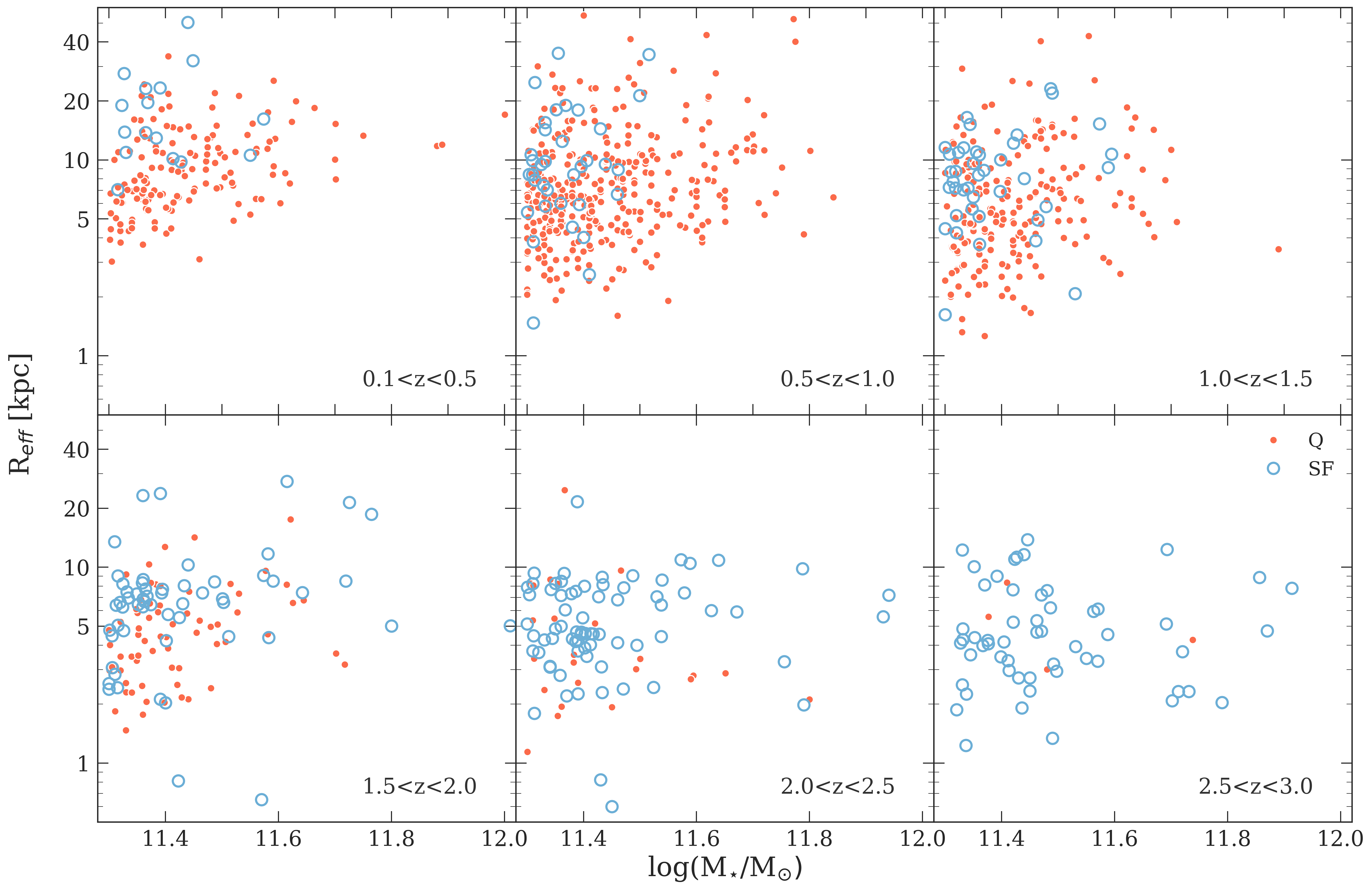}
\caption{Size-stellar mass distributions of star-forming and quiescent galaxies with $M_{\rm \star,corr}>2\times10^{11}M_{\odot}$ in six redshift bins. Star-forming galaxies are shown in blue circles and quiescent galaxies in red dot.} 
\label{fig:size_mass_dash_massive}
\end{figure*}

In order to estimate the uncertainty in the measured size of an individual galaxy, we place the best-fitting S\`ersic models in empty regions of our image mosaics, and refit them with GALFIT using the exact same procedures as described above. For each galaxy we create 100 realizations, using randomly chosen regions in the image mosaics. The uncertainties in $r_{\rm e}$ are then taken as the 16$^{\rm th}$ and
84$^{\rm th}$ values in the ordered list of 100 effective radii. This procedure ensures that the exact noise properties of the mosaics are taken into account.


\section{Sizes of the most massive galaxies}


The sizes of the most massive galaxies with $\log(M_{\rm \star,corr}/M_{\odot})>$11.3, as a function of stellar mass in six bins of redshift between 0.1$<$z$<$3.0, are presented in Figure \ref{fig:size_mass_dash_massive}. They blue circles represent the star-forming galaxies whereas the red dots represent the quiescent galaxies. Visually there is little difference between the size-mass distribution of the two populations, particularly at high redshift. In the following we quantify these relations and the offsets between star forming and quiescent galaxies.


\subsection{Median size and scatter of galaxies}

\begin{table*}[t]
\centering
\caption{Median sizes of galaxies as a function of galaxy mass and redshift. The redshift dependencies, parameterized by $r_{\rm eff} = B_z \times (1+z)^{-\beta_z}$, are also given.}
\begin{tabular}{|c|cc|cc|cc}
\hline
\multirow{2}{*}{z} & \multicolumn{2}{c|}{\textbf{All}} & \multicolumn{2}{c|}{\textbf{Star-forming}} & \multicolumn{2}{c|}{\textbf{Quiescent}} \\ \cline{2-7} 
 & Med r$_{\rm eff}$ (kpc) & $\sigma$ $\log$ r$_{\rm eff}$ &Med r$_{\rm eff}$ (kpc)  & $\sigma$ $\log$ r$_{\rm eff}$ & Med r$_{\rm eff}$ (kpc)  & \multicolumn{1}{c|}{$\sigma$ $\log$ r$_{\rm eff}$} \\ \hline
0.25 & 9.34 $\pm$0.34 & 0.22$\pm$0.01 & 15.74$\pm$2.21 & 0.21$\pm$0.01 & 8.84$\pm$0.36 & \multicolumn{1}{c|}{0.22$\pm$0.04} \\
0.75 & 6.99$\pm$0.20 & 0.27$\pm$0.01  & 9.25$\pm$1.0 & \multicolumn{1}{c|}{0.30$\pm$0.04} & 6.76$\pm$0.20 & 0.26$\pm$0.01\\
1.25 & 6.26$\pm$0.26 & 0.28$\pm$0.01  & 8.33$\pm$0.66 & \multicolumn{1}{c|}{0.22$\pm$0.03} & 5.71$\pm$0.26 & 0.28$\pm$0.01\\
1.75 & 5.33$\pm$0.25 & 0.27$\pm$0.02 & 6.33$\pm$0.37 & \multicolumn{1}{c|}{0.23$\pm$0.02} & 4.38$\pm$0.29 & 0.23$\pm$0.02 \\
2.25 & 5.02$\pm$0.28 & 0.26$\pm$0.02 &  5.34$\pm$0.32 & \multicolumn{1}{c|}{0.23$\pm$0.02} & 3.56$\pm$0.54 & 0.30$\pm$0.04 \\
2.75 & 4.32$\pm$0.38 & 0.27$\pm$0.03 & 4.32$\pm$0.40 & \multicolumn{1}{c|}{0.27$\pm$0.02} & 5.07$\pm$1.22 & 0.17$\pm$0.06 \\ \hline
B$_z$ & \multicolumn{2}{c|}{10.4$\pm$0.4} & \multicolumn{2}{c|}{18.0$\pm$2.1} & \multicolumn{2}{c|}{10.8$\pm$0.4} \\
$\beta_z$ & \multicolumn{2}{c|}{0.65$\pm$0.05} & \multicolumn{2}{c|}{1.04$\pm$0.11} & \multicolumn{2}{c|}{0.84$\pm$0.06} \\ \hline
\end{tabular}
\label{tab:med}
\end{table*}

\begin{figure*}[]
\centering
\includegraphics[width=\textwidth]{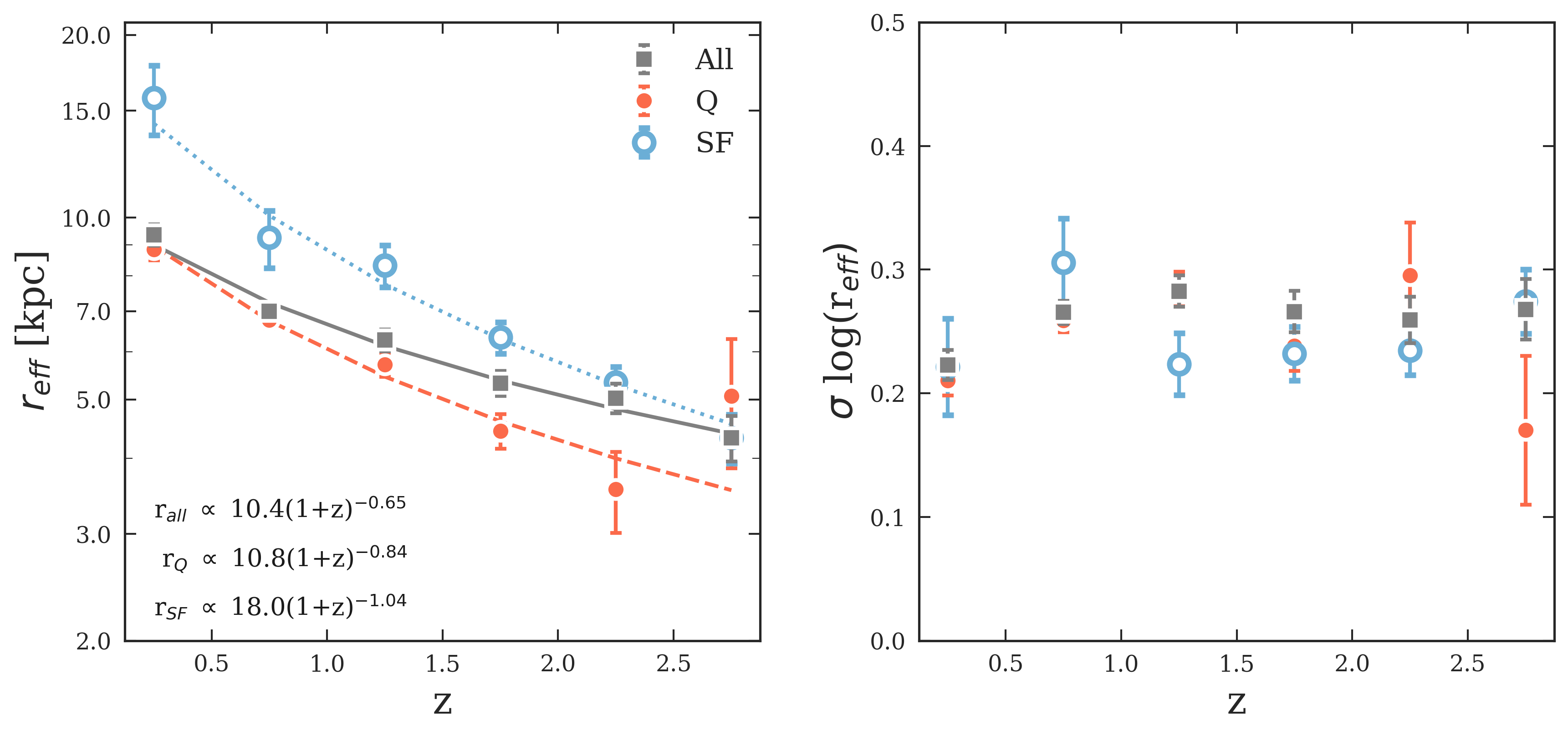}
\caption{Median size (left) and observed scatter in size (right) of massive galaxies as a function of redshift for all galaxies (grey squares), quiescent galaxies (red dots) and star-forming galaxies (blue circles). In the left panel, the lines represent the fits to the median sizes of the form r$_{\rm eff}/$kpc $=$ B$_z$(1+z)$^{\beta_z}$. The median size of star-forming galaxies is larger and evolves slightly more rapidly with redshift than that of the quiescent galaxies. The median size of all galaxies is similar to that of star-forming galaxies at higher redshift and to quiescent galaxies at lower redshift, with the overall evolution being shallower than either of the populations. The right panel shows the evolution of the observed scatter in the sizes of the different populations of galaxies. Within errors, there is no difference in the observed scatter between the populations. The median sizes and observed scatter of all, star-forming and quiescent galaxies are given in Table \ref{tab:med}. } 
\label{fig:med_scatter}
\end{figure*}

We first examine the distribution of stellar mass within our mass-limited sample. The median stellar mass of all galaxies with $\log(M_{\star}/M_{\odot})>$ 11.3 at z$\sim$0.25 is $\log(M_{\star}/M_{\odot})=$ 11.43$\pm$0.01. At this epoch quiescent and star-forming galaxies have a similar median stellar mass. At $z\sim2.75$, the median stellar mass of all galaxies is $\log(M_{\star}/M_{\odot})=$ 11.46$\pm$0.03. The star-forming  galaxies have a similar median stellar mass as the full sample as they make up the bulk of the population at this epoch; the few quiescent galaxies are slightly less massive with $\log(M_{\star}/M_{\odot})=$ 11.36$\pm$0.04. 

Next, we use the biweight estimator for the location and scale of a distribution to calculate the median size and its scatter,  as this estimator gives higher weights to the center of distribution and is insensitive to outliers.

Figure \ref{fig:med_scatter} shows the evolution of the median size of galaxies in 0.1$<$z$<$3.0. The median size of all galaxies and of the two sub-populations have increased from z$\sim$3 to z$\sim$0.1. The median size of all galaxies at 2.5$<$z$<$3.0 was 4.3$\pm$0.4 kpc which has increased to 9.3$\pm$0.4 kpc  at 0.1$<$z$<$0.5. The median size of quiescent galaxies at 2.5$<$z$<$3.0 was 5.1$\pm$1.2 kpc which has increased to 8.8$\pm$0.4 kpc  at 0.1$<$z$<$0.5.  The median size of star-forming galaxies at 2.5$<$z$<$3.0 was 4.3$\pm$0.4 kpc which has rapidly increased to 15.7$\pm$2.1 kpc at 0.1$<$z$<$0.5. Note that the median size of all galaxies changes more slowly with redshift than that of either of the subpopulation; this reflects of the fact that the fraction of quiescent galaxies increases from 13 $\%$ at 2.5$<$z$<$3.0 to 90 $\%$ 0.1$<$z$<$0.5. This is discussed in more detail in Section \ref{sec:transition}. We determine the significance of the difference in the sizes of quiescent and star-forming galaxies with the two-sided Mann-Whitney test. The probability that the quiescent and star-forming galaxies are drawn from the same sample at 0.1$<$z$<$0.5 is less than $5 \times 10^{-5}$. This probability rises to more than 40 $\%$ at 2.5$<$z$<$3.0, and we conclude that the sizes of the most massive star forming and quiescent galaxies are not significantly different at $2.5<z<3.0$. However, due to the small sample of Q galaxies at $z>2.5$ we cannot rule out that a small size differences emerges when more data are available.

We also present the measurement of the observed scatter in the sizes for all massive galaxies and the subsamples of quiescent and  star forming galaxies. We find no strong evolution in the observed scatter for the full sample, or either of the subpopulations. The scatters range between $0.2-0.3$\,dex, in good agreement with the values found by previous studies.

\subsection{Evolution of the median size}

We parameterize the evolution of the median size of the galaxies as
\begin{equation}
    r_{\rm eff} = B_z \times (1+z)^{-\beta_z}.
\end{equation}
The grey, red and blue lines in Figure \ref{fig:med_scatter} represent the evolution of the median sizes of all galaxies, the quiescent galaxies, and the star-forming galaxies respectively.
We find $\beta=$0.65$\pm$0.05 for the full sample.
For the star-forming galaxies $\beta_z=$1.04$\pm$0.11, and for quiescent galaxies we find $\beta=$0.84$\pm$0.06. The size evolution of the star-forming and the quiescent galaxies is only marginally significant. This is consistent with the results of \citet{Faisst2017} who  found a similar evolution for ultra-massive star-forming and quiescent galaxies at 0.5$<$z$<$2.5 (although they found a slightly faster evolution of 1.21$\pm$0.20 for the quiescent population). This behavior is qualitatively different from the evolution of lower mass galaxies, where quiescent galaxies show a more rapid size evolution than star-forming galaxies. Specifically, \citet{VanderWel2014a} find that galaxies with $\log(M_{\star}/M_{\odot}\sim 10.75$ have $\beta_z=$1.24$\pm$0.08 and 0.72$\pm$0.09 for quiescent and star-forming galaxies respectively.

\section{Evolution of the size-mass distribution from}
\label{sec:evolution}

\begin{figure*}[t]
\centering
\includegraphics[width=\textwidth]{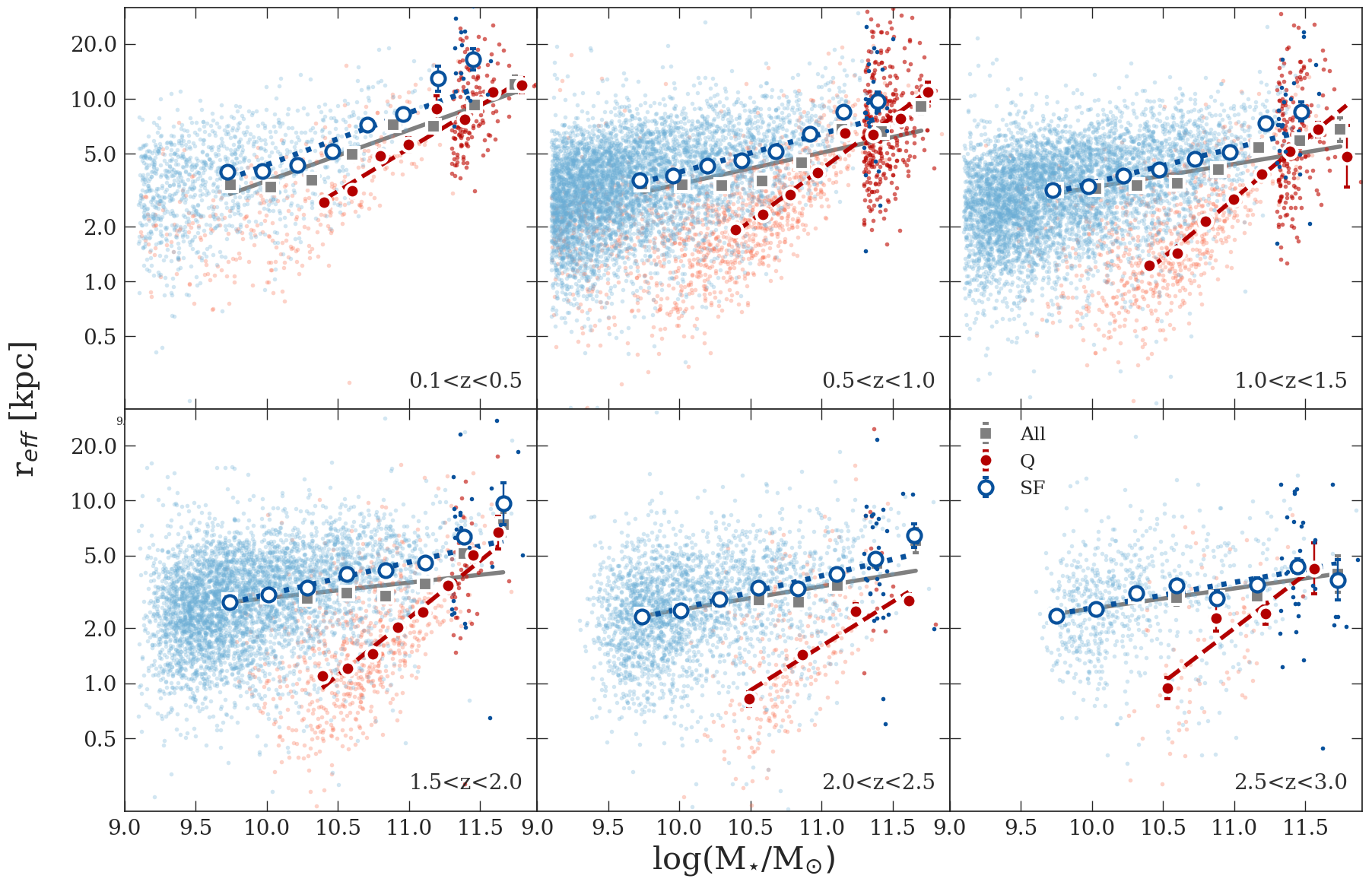}
\caption{Size-stellar mass distribution of star-forming and quiescent galaxies for $\log(M_{\star}/M_{\odot}>$9.0 at $0.1<z<3.0$. The small dots in the background show the combined sample of \citep{VanderWel2014a} and the most massive galaxies from this study. The points show the median sizes of all the galaxies with their $1\sigma$ dispersion. The lines indicate fits to the size-mass relation of all, star-forming, and quiescent galaxies. The median sizes of intermediate mass star-forming and quiescent galaxies are significantly different at a given stellar mass at all redshifts. However, at the high mass end the gap closes with the two populations having similar sizes, in agreement with previous studies.} 
\label{fig:size_all}
\end{figure*}

\begin{figure*}[t]
\centering
\includegraphics[width=\textwidth]{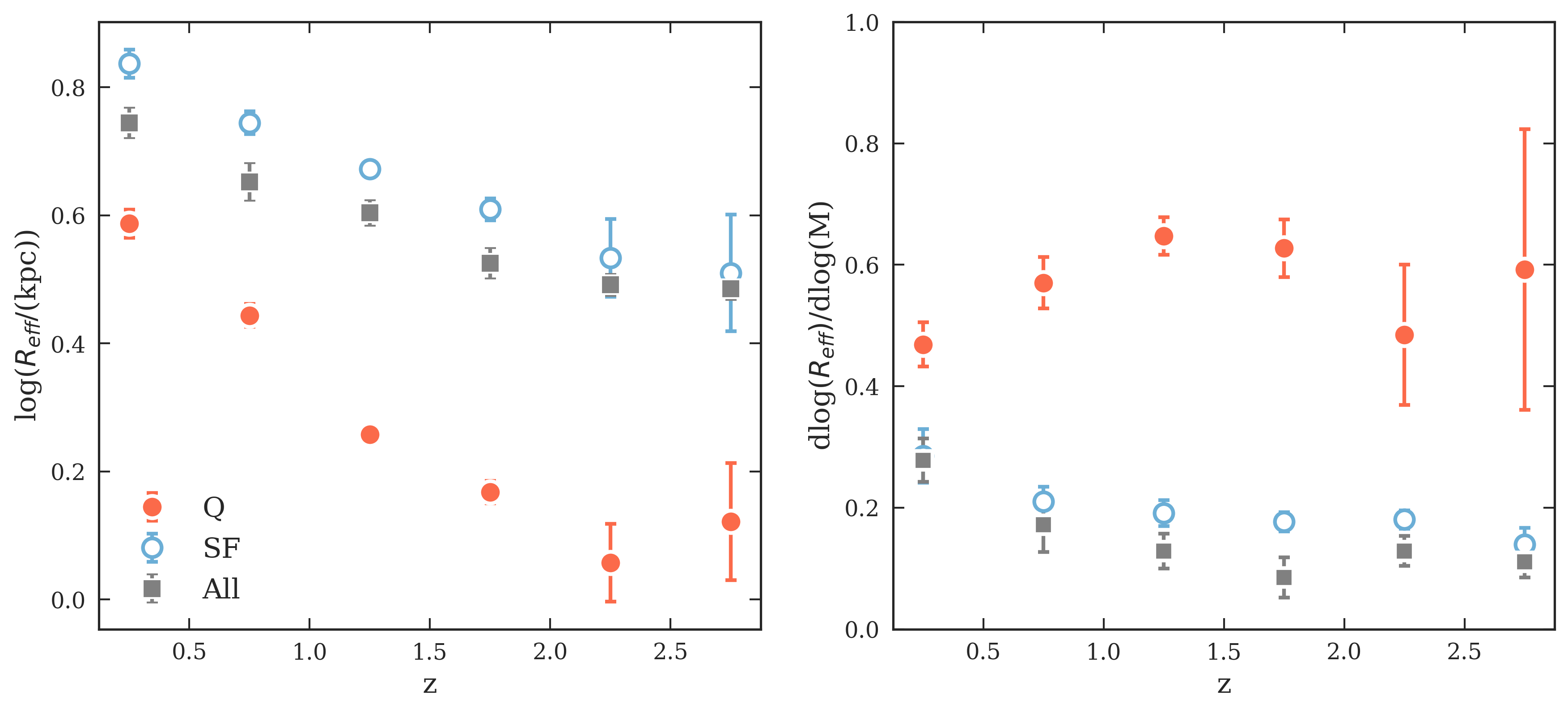}
\caption{Redshift evolution of the parameters of the analytic size mass relations. The left panel shows the intercept and the right panel shows the slope of the power law fits shown in Figure \ref{fig:size_fit} at fixed stellar mass of $M_{\star}=5\times10^{10}M_{\odot}$.}
\label{fig:size_fit}
\end{figure*}

In this Section we combine our size measurements of the most massive galaxies with the large \citet{VanderWel2014a} sample to study the size-mass distribution over a large dynamical range of mass and redshift. Figure \ref{fig:size_all} shows the size-mass distribution of galaxies with $\log(M_{\star}/M_{\odot})>$9.0 at 0.1$<$z$<$3.0. 
Visually, the most massive galaxies follow extrapolations from the trends seen for less massive galaxies. 
We fit the size-mass distribution of galaxies with $\log(M_{\star}/M_{\odot})>$9.0 at 0.1$<$z$<$3.0. We fit the star-forming and quiescent populations separately as has been done previously by \citet{VanderWel2014a}, but now with a more populated high mass end. We also provide fits to the full sample of galaxies, that is, not split up by their stellar populations.

\subsection{Analytic Fits to Galaxy Size Evolution}

\begin{table*}[]
\centering
\caption{Results from the parameterized fits to the size-mass distribution of the form $r_{\rm eff}$/kpc $=$ A(M$_{\star}$/5 $\times$ 10$^{10}$ M$_{\odot}$)$^{\alpha}$, as described in Section \ref{sec:size} and shown in Figures \ref{fig:size_fit}}
\begin{tabular}{|c|ll|cc|cc|}
\hline
\multirow{2}{*}{z} & \multicolumn{2}{c|}{\textbf{All}} & \multicolumn{2}{c|}{\textbf{Star-forming}} & \multicolumn{2}{c|}{\textbf{Quiescent}} \\ \cline{2-7} 
 & $\log$ (A) & $\alpha$ & $\log$ (A) & $\alpha$ & $\log$ (A) & $\alpha$ \\ \hline
0.25 & 0.74$\pm$0.02 & 0.27$\pm$0.03 & 0.83 $\pm$ 0.02 & 0.29 $\pm$ 0.04 & 0.59 $\pm$ 0.02 & 0.47 $\pm$ 0.03 \\
0.75 & 0.65$\pm$0.03 & 0.17$\pm$0.04 & 0.74 $\pm$ 0.02 & 0.21 $\pm$ 0.02 & 0.44 $\pm$ 0.02 & 0.57$\pm$ 0.04 \\
1.25 & 0.60$\pm$0.02 & 0.13$\pm$0.03 & 0.67 $\pm$ 0.01 & 0.19 $\pm$ 0.02 & 0.26 $\pm$ 0.01 & 0.65 $\pm$ 0.03 \\
1.75 & 0.53$\pm$0.02 & 0.09$\pm$0.03 & 0.61 $\pm$ 0.01 & 0.18 $\pm$ 0.01 & 0.17$\pm$ 0.02 & 0.63 $\pm$ 0.05 \\
2.25 & 0.49$\pm$0.01 & 0.13$\pm$0.02 & 0.53 $\pm$ 0.06 & 0.18 $\pm$ 0.01 & 0.06 $\pm$ 0.06 & 0.48 $\pm$ 0.11 \\
2.75 & 0.48$\pm$0.02 & 0.11$\pm$0.03 & 0.51 $\pm$ 0.09 & 0.14 $\pm$ 0.03 & 0.12 $\pm$ 0.09 & 0.59 $\pm$ 0.23 \\ \hline
\end{tabular}
\label{tab:fit}
\end{table*}

We  fit the size-mass relation of the combined sample of the most massive galaxies of this study and the \citet{VanderWel2014a} sample. Following \citet{Shen2003TheSurvey}, we assume a log-normal distribution N($\log (r_{\rm eff})$, $\sigma_{\log r_{\rm eff}}$, where $\log (r_{\rm eff})$ is the mean and $\sigma_{\log r_{\rm eff}}$ is the dispersion. Similar to \citet{VanderWel2014a}, $r_{\rm eff}$ is taken to be a function of galaxy mass such that:

\begin{equation}
    r_{\rm eff}(m_{\star})/{\rm kpc} = A \times m_{\star}^{\alpha},
\end{equation}

where $m_{\star}$ $\equiv$ M$_{\star}$/7 $\times$ 10$^{10}$ M$_{\odot}$. The basic characteristics of the galaxy size distribution are given by the slope $\alpha$, intercept $A$, and intrinsic scatter $\sigma_{\log r}$ of size as a function of mass. Following \citet{VanderWel2014a}\, we fit all star-forming galaxies with M$_{\star}$ $>$ 3 $\times$ 10$^9$ M$_{\odot}$ and for the quiescent sample we fit galaxies with M$_{\star}$ $>$ 2 $\times$ 10$^{10}$ M$_{\odot}$, in the redshift range $0.1<z<3.0$. This stellar mass limit is derived from integrated magnitude limit, as GALFIT is able to reasonably estimate effective radius for galaxies with m$<$24.0 and S\`ersic index for galaxies with m$<$23.

The analytic fits to the size-mass distributions are shown in Figure \ref{fig:size_all}.  The lines in Figure \ref{fig:size_all} indicate the best-fit relations, while the evolution of the individual parameters (intercept $A$ and slope $\alpha$) are shown in Figure \ref{fig:size_fit}. The best-fit parameters are also given in Table \ref{tab:fit}. These new analytic fits have been performed on the same data for stellar mass $\log$(M$_{\star}$/M$_{\odot})<$ 11.3 as of \citet{VanderWel2014a} but on a three times larger dataset for $\log$(M$_{\star}$/M$_{\odot})>$ 11.3 at z$>$1.5 and on a five times larger dataset at z$<$1.5. Our size-mass relation of star-forming galaxies agrees well with that found by the \citet{VanderWel2014a}; in other words, the sizes of the most massive star-forming galaxies are similar to those expected from their relations. The slope of the size-mass relation of star-forming galaxies is $\approx$0.25, with the slope slightly decreasing with redshift. However, we find a shallower size-mass relation of quiescent galaxies than \citet{VanderWel2014a}; we find an approximate constant slope at all redshifts of $0.5-0.6$, whereas \citet{VanderWel2014a} found a slope of $\approx 0.75$ (at all redshifts). 

Finally, we fit the redshift evolution of the parameters of the best-fitting relations, to arrive at a complete description of the sizes of galaxies as a function of mass and redshift. For the full sample of galaxies the parameters evolve as
$\log A=-0.26$ $\log (1+z)+0.60$ and $\alpha=-0.17\log (1+z)+0.16$. For the star forming galaxies we find $\log A = -0.32\log(1+z)+0.67$ and $\alpha=-0.12\log(1+z)+0.21$. For the quiescent galaxies the slope does not evolve significantly with redshift, and we derive $\log A=-0.52\log(1+z)+0.31$ and $\alpha=0.57$.

\subsection{Evolution of the Median Sizes of Galaxies}

\begin{figure*}[t]
\centering
\includegraphics[width=0.98\textwidth]{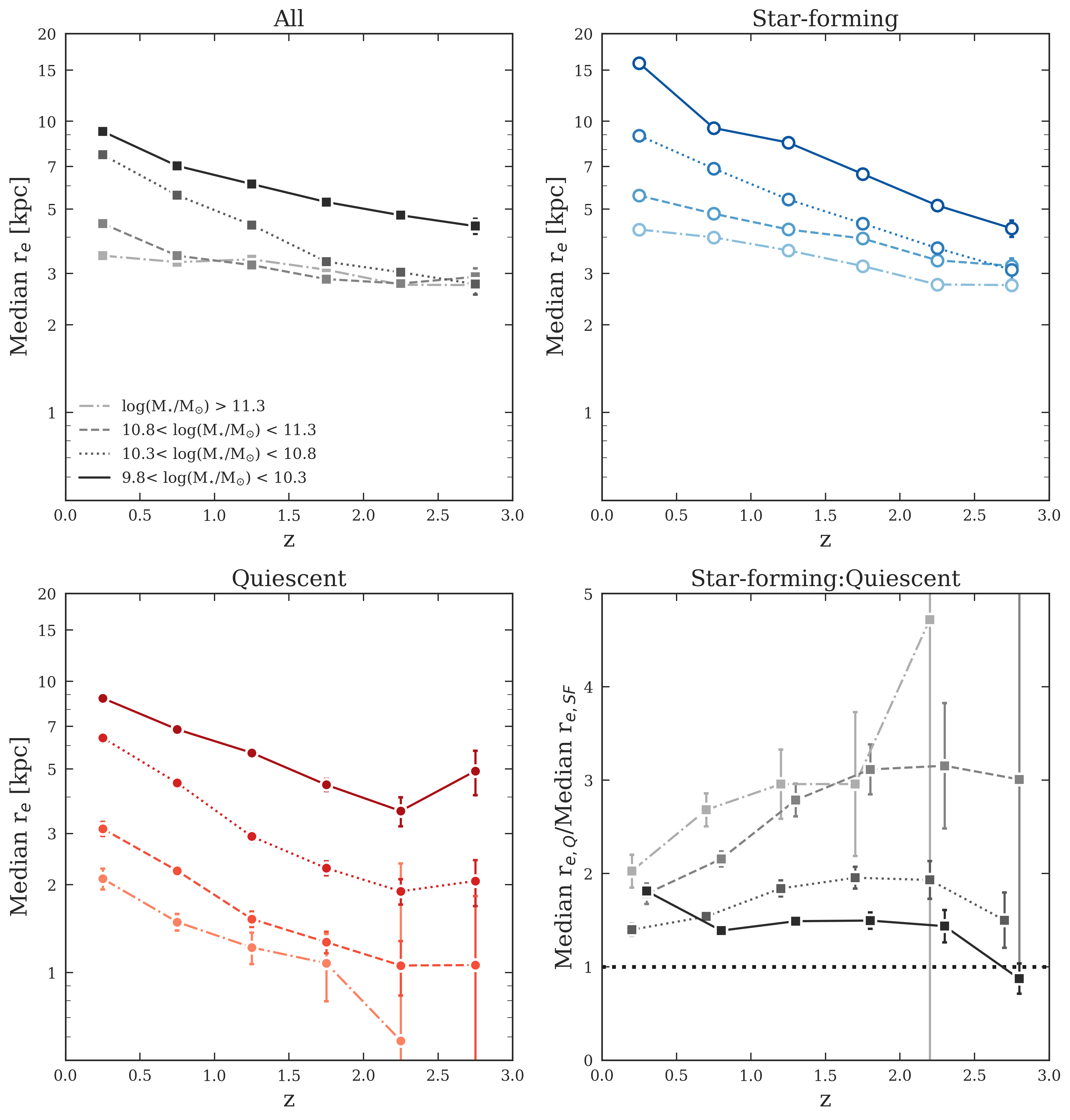}
\caption{Median size of all (top-left), star-forming (top-right) and quiescent (bottom-left) galaxies, and ratio of median sizes of star-forming and quiescent galaxies (bottom-right) as a function of stellar mass and redshift. The median sizes are calculated from the combined sample of this study and \citet{VanderWel2014a}. } 
\label{fig:size_med}
\end{figure*}

So far we have considered the relation between size and mass at fixed redshifts. In Fig.\ \ref{fig:size_med} we show the complementary information, that is, the relation between size and redshift for particular masses. The median sizes are calculated from the combined sample of this study and \citet{VanderWel2014a}. The trends in this Figure show the same behavior as discussed earlier in the paper.
The slope of the evolution of the full sample does not change significantly with stellar mass. The evolution of quiescent galaxies is not strongly mass-dependent and is more rapid than that of star forming  galaxies. The rate of evolution of star-forming galaxies increases with stellar mass, such that the redshift evolution of the most massive star forming galaxies is very similar to that of the most massive quiescent galaxies. 

\begin{table*}[]
\centering
\caption{Median sizes of galaxies as a function of stellar mass and redshift for all, star-forming and quiescent galaxies. The stellar masses in the header and the redshifts in the first column are the centers of each 0.5-wide bins. Evolution of size with redshift for each stellar mass bin in the form of r$_{\rm eff}$/kpc $=$ B$_{\rm z}(1+z)^{-\beta_z}$ are also given.}
\begin{tabular}{|c|cccc||cccc||cccc|}
\hline
\multirow{2}{*}{\textbf{z}} & \multicolumn{4}{c||}{\textbf{ All} Median \textbf{r$_e$ (kpc)}} & \multicolumn{4}{c||}{\textbf{Star-forming} Median \textbf{r$_e$ (kpc)}} & \multicolumn{4}{c|}{\textbf{Quiescent} Median \textbf{r$_e$ (kpc)}} \\ 
 & \multicolumn{1}{c|}{$M_{\star}=10^{10}$} & \multicolumn{1}{c|}{10.5} & \multicolumn{1}{c|}{11} & \multicolumn{1}{c||}{$>$11.3} & \multicolumn{1}{c|}{$M_{\star}=10^{10}$} & \multicolumn{1}{c|}{10.5} & \multicolumn{1}{c|}{11} & \multicolumn{1}{c||}{$>$11.3}& \multicolumn{1}{c|}{$M_{\star}=10^{10}$} & \multicolumn{1}{c|}{10.5} & \multicolumn{1}{c|}{11} & \multicolumn{1}{c||}{$>$11.3} \\ \hline
0.25 & 3.5$\pm$0.1 & 4.5$\pm$0.1 & 7.7$\pm$0.2 & 9.2$\pm$0.1  & 4.2$\pm$0.1 & 5.6$\pm$0.1 & 8.9$\pm$0.3 & 15.8$\pm$0.4 & 2.1$\pm$0.2 & 3.1$\pm$0.2 & 6.4$\pm$0.2 & 8.7$\pm$0.1 \\
0.75 & 3.3$\pm$0.1 & 3.5$\pm$0.1 & 5.6$\pm$0.1 & 7.0$\pm$0.1 & 4.0$\pm$0 & 4.8$\pm$0.1 & 6.9$\pm$0.1 & 9.5$\pm$0.3 & 1.5$\pm$0.1 & 2.2$\pm$0.1 & 4.5$\pm$0.1 & 6.8$\pm$0.1 \\
1.25 & 3.4$\pm$0 & 3.2$\pm$0.1 & 4.4$\pm$0.1 & 6.1$\pm$0.1  & 3.6$\pm$0 & 4.3$\pm$0.1 & 5.4$\pm$0.1 & 8.5$\pm$0.3 & 1.2$\pm$0.2 & 1.5$\pm$0.1 & 2.9$\pm$0.1 & 5.7$\pm$0.1\\
1.75 & 3.1$\pm$0.1 & 2.9$\pm$0.1 & 3.3$\pm$0.1 & 5.3$\pm$0.2 & 3.2$\pm$0.1 & 4.0$\pm$0.1 & 4.5$\pm$0.1 & 6.6$\pm$0.2& 1.1$\pm$0.3 & 1.3$\pm$0.1 & 2.3$\pm$0.1 & 4.4$\pm$0.2  \\
2.25 & 2.7$\pm$0.1 & 2.8$\pm$0.1 & 3.0$\pm$0.1 & 4.8$\pm$0.2 & 2.8$\pm$0.1 & 3.3$\pm$0.1 & 3.7$\pm$0.1 & 5.1$\pm$0.2 & 0.6$\pm$1.8 & 1.1$\pm$0.2 & 1.9$\pm$0.2 & 3.6$\pm$0.4 \\
2.75 & 2.7$\pm$0.1 & 2.9$\pm$0.2 & 2.8$\pm$0.2 & 4.4$\pm$0.3 & 2.7$\pm$0.1 & 3.2$\pm$0.2 & 3.1$\pm$0.3 & 4.3$\pm$0.3 & ... & 1.1$\pm$0.8 & 2.1$\pm$0.4 & 4.9$\pm$0.9 \\
\hline \hline
\multicolumn{1}{|l|}{B$_z$} & \multicolumn{1}{c}{3.7$\pm$0.2} & \multicolumn{1}{l}{4.6$\pm$0.2} & 9.7$\pm$0.3 & 10.7$\pm$0.2 & \multicolumn{1}{c}{5.1$\pm$0.2} & \multicolumn{1}{l}{6.3$\pm$0.2} & 11.0$\pm$0.3 & 19.8$\pm$1.2 & \multicolumn{1}{c}{2.5$\pm$0.1} & \multicolumn{1}{l}{4.1$\pm$0.2} & 8.8$\pm$0.5 & 10.5$\pm$0.3 \\
\multicolumn{1}{|l|}{$\beta_z$} & \multicolumn{1}{c}{0.2$\pm$0.1} & \multicolumn{1}{l}{0.2$\pm$0.1} & 1.0$\pm$0.1 & 0.7$\pm$0.1 & \multicolumn{1}{c}{0.4$\pm$0.1} & \multicolumn{1}{l}{0.5$\pm$0.1} & 0.9$\pm$0.03 & 1.1$\pm$0.1 & \multicolumn{1}{c}{0.9$\pm$0.1} & \multicolumn{1}{l}{1.1$\pm$0.1} & 1.3$\pm$0.1 & 0.8$\pm$0.1 \\ \hline

\end{tabular}
\end{table*}

\section{Discussion}
\label{sec:discussion}

\subsection{Comparison to Previous Studies}

We present the first comprehensive measurements of the sizes of the most massive galaxies with $M_{\star}>2\times 10^{11}$ M$_{\odot}$ within $0.1<z<3.0$ measured at HST resolution. As shown in Figure \ref{fig:size_med}, we confirm that the galaxies in this study are larger in size than the less massive galaxies \citep{Shen2003TheSurvey, Carollo2013,VanderWel2014a} at the same epoch. The bottom-left panel of Figure \ref{fig:size_med} shows the ratio of median sizes of star-forming galaxies to that of quiescent galaxies. For intermediate to massive galaxies, with stellar masses between $ 10^{9}$ and $10^{11} M_{\odot}$, quiescent galaxies are, on average, smaller than star-forming galaxies \citep{Corollo2013,VanderWel2014a}. At z$\sim$2.25, for stellar mass of $M_{\star}\sim 5 \times 10^{10}$ M$_{\odot}$ the median size of star-forming galaxy is 3.39$\pm$0.08 kpc and the median size of quiescent galaxy is 1.20$\pm$0.03 kpc, which is almost a third of the size of the star-forming galaxies. However, for galaxies with stellar masses $\geq 2 \times 10^{11} M_{\odot}$ the two classes of galaxies are found to have similar sizes of 5.1$^{+0.6}_{-0.1}$ kpc and 3.6$^{+1.9}_{-0.4}$ kpc. 

Our results are consistent with the measurements of \citealt{VanderWel2014a} in the same mass range, although their sample is smaller by a factor of 3--4. More precisely, they confirm the extrapolated size-mass relations of \citet{VanderWel2014a}, which were dominated by less massive galaxies.
Our results are also in agreement with the ground-based measurements of \citet{Faisst2017} and \citet{Hill2017}. \citet{Faisst2017} studied galaxies with $\log(M_{\star}/M_{\odot})>11.4$, slightly more massive than galaxies in our sample. Within the uncertainties, the sizes of these galaxies are fully consistent with our median sizes.

\subsection{Central Density of the Most Massive Galaxies}
\label{sec:transition}

\begin{figure}[ht]
\centering
\includegraphics[width=0.46\textwidth]{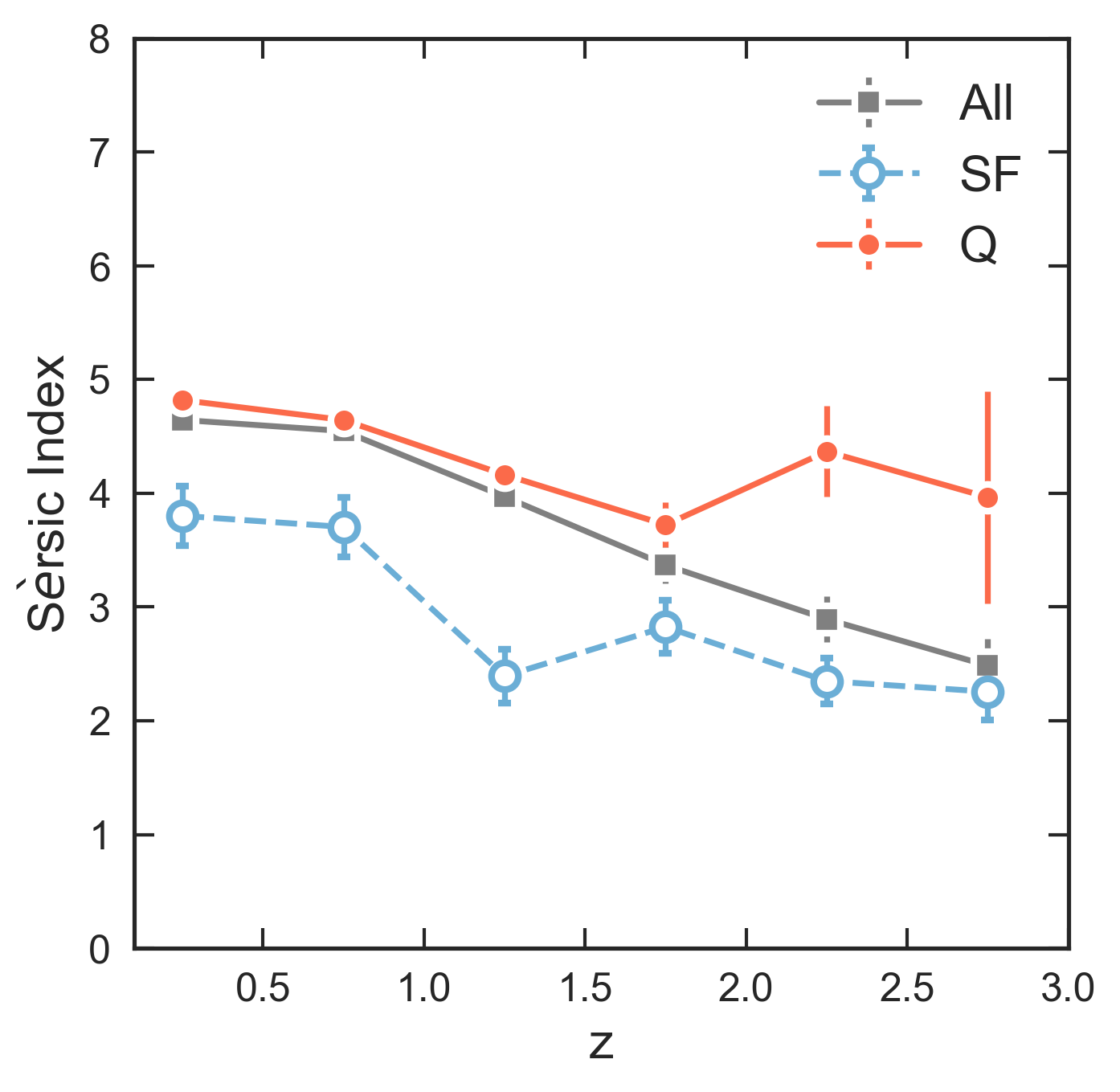}
\caption{Evolution of median S\`ersic indices with redshift of all, star-forming and quiescent galaxies.}
\label{fig:Sersic}
\end{figure}

\begin{figure}[ht]
\centering
\includegraphics[width=0.5\textwidth]{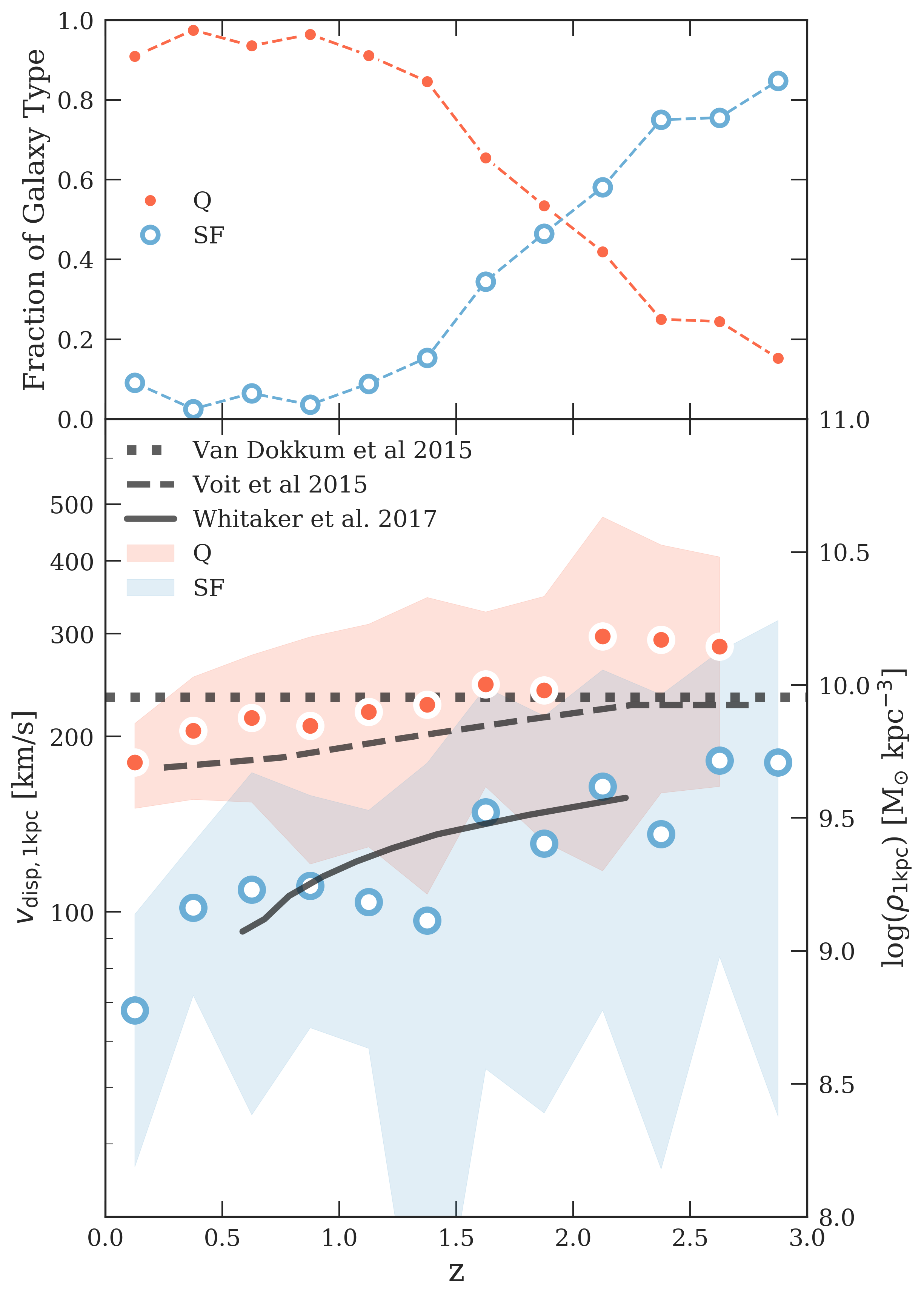}
\caption{Top panel: Evolution of the fraction of the most massive star forming and quiescent galaxies. The fraction of star-forming galaxies increases with redshift while that of quiescent galaxies decreases with the cross-over happening at z$\sim$2. Bottom panel: Evolution of the central mass density of star-forming and quiescent galaxies. The central mass density is the density of the central 1 kpc of the galaxy, calculated using Eq.\ \ref{eq:vsig}. The blue circles and red dots show the median central densities of the star-forming and quiescent populations, and the hatched areas show 1-$\sigma$ spread. The black line is the median velocity dispersion of all quenched galaxies with $9.0<\log(M_{\star}/M_{\odot})<12.0$ measured by \citet{Whitaker2015}. The dotted black line is the assumed constant threshold in velocity dispersion above which galaxies at $1.5<z<3.0$ quench from \citet{VanDokkum2015_compactmassive}. The dashed black line is the predicted quenching threshold from \citet{Voit2015Precipitation-RegulatedGalaxies} normalized to 300 km/s at z$=$2. } 
\label{fig:core}
\end{figure}

Studies of the central galaxy density out to z$=$3 \citep{Cheung2012The0.5,Saracco2012OnGalaxies,Barro2012CANDELS:Z2,Barro2015} find that the innermost structure of galaxies correlates with the star formation rate. Quenching of galaxies (whether by AGN feedback or other mechanisms) is thought to become very efficient when the central density reaches a certain threshold \citep{VanDokkum2015_compactmassive,Whitaker2016}.    As stellar density and velocity dispersion are closely related, observations therefore indicate that galaxies are statistically more likely to be quiescent once they have surpassed a threshold in either density or velocity dispersion. This has been studied in detail by \citet{Whitaker2016} who found an abrupt cessation of star formation when galaxies reach a threshold central stellar density.

We can use the information of the two dimensional light profiles of the galaxies and their total stellar mass to infer the central stellar density of the galaxy, assuming spherical symmetry. We follow the prescription from \citet{Bezanson2009} and \citet{Whitaker2016} to calculate the central stellar density and central velocity dispersions of the galaxies. We perform an Abel transform to deproject the circularized, three-dimensional light profile using:

\begin{equation}
    \rho(x) = \frac{b_n}{\pi} \frac{I_0}{r_{\rm eff}} x^{1/n-1} \int^{\infty}_1 \frac{{\rm exp}(-b_nx^{1/n}t)}{\sqrt{t^{2n}-1}} dt,
\end{equation}

with $\rho$ the three-dimensional (3D) luminosity density in a particular filter, $x\equiv r/r_{\rm eff}$, $r_{\rm eff}$ the circularized effective radius, the $n$ the S\`ersic index, and $b_n$ the $n$-dependent normalization parameter of the S\`ersic profile. We note that this methodology may lead to errors for galaxies that are far from spherical symmetry, in particular for flat disks. The mass within $r=1$\,kpc is calculated by integrating the 3D luminosity profiles.
Assuming mass follows light and neglecting color gradients, we convert the central luminosity to central stellar mass using the corrected total stellar masses from the UVISTA catalog. The central density is calculated by numerically integrating the following equation:

\begin{equation}
    \rho_{\rm 1kpc} = \frac{\int^{\rm 1kpc}_0 \rho(r)r^2 dr }{\int^{\infty}_0 \rho(r)r^2 dr}\frac{M_{\rm tot}}{\frac{4}{3}\pi \rm(1 kpc)^3}.
\label{eq:vsig}
\end{equation}

Figure \ref{fig:core} shows the central stellar densities of star-forming and quiescent galaxies in our sample. The central velocity dispersion is calculated assuming $v_{\rm disp, 1kpc}=v_{\rm circ, 1kpc}/\sqrt{2}$, where $v_{\rm circ, 1kpc}=\sqrt{4/3 G \rho_{\rm 1kpc}}$ and $G$ is the gravitational constant. As can be seen, the central stellar density of both populations of galaxies are lower at $z=0$ than at $z\sim3$, but the central density of quiescent galaxies is always higher than that of star-forming galaxies: the equivalent velocity dispersion of star-forming galaxies  decreased from 181 km/s to 67 km/s whereas that of quiescent galaxies decreased from 285 km/s to 180 km/s between z$\sim$2.63 and z$\sim$0.13.

This difference in central density between star forming and quiescent galaxies may seem surprising, given the fact that they have very similar sizes at fixed mass. However, the Sersic indices also enter the calculation of the central densities, and they are significantly different between the two samples: we find $\langle n\rangle=4.0 \pm 0.$ for quiescent galaxies at $z\sim 2.75$ and $\langle n\rangle=2.2 \pm 0.2$ for star forming galaxies, as shown in Fig. \ref{fig:Sersic}. The difference in central density between the two samples is driven by this difference in the profile shape.

We show quenching threshold velocity dispersions from previous studies in Fig.\ \ref{fig:core}. As can be seen the median central density of quenched galaxies are always  above the threshold density while that of star-forming galaxies is above the threshold density only at the highest redshifts.
This analysis thus provides evidence that it is indeed the central density, rather than the average surface density within the effective radius \citep[e.g.,][]{Franx2008,Maier2009The0.5}, that determines whether a galaxy is star forming or quenched.
We note here that quenching may not be a single event in the life of a galaxy; specifically, many of the massive star forming galaxies at low redshift may be rejuvenated by fresh gas infall. These transitions change the fraction of quiescent and star forming galaxies, with young quiescent galaxies being added to the sample (Fig.\ \ref{fig:core} upper panel, also see, e.g., \citet{VanDokkum2001_morphology, Carollo2013}).

\section{Summary}
\label{sec:conclusion}

In this paper we presented the COSMOS-DASH mosaic, the widest-area near-IR imaging survey yet done with HST. We used this dataset to measure the distribution of \totgal\ galaxies with $M_{\star}>2\times10^{11}M_\odot$ in the size-mass plane over the redshift range 0.1$<$z$<$3.0. We also combine this sample with the extensive sample of lower mass galaxies of \citet{VanderWel2014a}. We find the following:
\begin{itemize}
    \item 
We find that the size of galaxies increases with their mass, and that this trend continues at the highest masses.
Some intriguing individual exceptions exist: we find a small number of extremely small and massive galaxies and it will be interesting to obtain follow-up spectroscopy of these objects \citep[see, e.g.,][]{Nelson2014}. 
\item
The size of the most massive galaxies decreases with increasing redshift.  As shown in Fig.\ \ref{fig:med_scatter} the size of the most massive star-forming galaxies decreases by a factor of $\approx 3.5$ from $z=0.2$ to $z=2.75$, and the most massive guiescent galaxies decrease in size by a factor of $\approx 2.5$ over that same redshift range. 
\item
The evolution of the {\em ratio} between the sizes of star forming galaxies and quiescent galaxies shows a strong mass-dependence. At low to intermediate masses, the size difference between these populations increases with redshift; however, at the highest masses quiescent and star forming galaxies are approximately the same size at all redshifts (see Fig.\ 13).
\item
We derive analytic fits to the relations between mass and size, and between the parameters of these fits with redshift. Together this set of equations provides a complete description of the median sizes of galaxies as a function of mass, redshift, and star formation rate. 
The data presented hear are consistent with most published data sets, although very few have focused specifically at this mass end.
\item
Finally, we inferred the central stellar density and velocity dispersion of the galaxies and show that the central densities of quiescent galaxies are higher than those of star-forming galaxies even though their sizes are similar. The fact that the most massive star forming galaxies at $z=2.5-3.0$ are close to the ``quenching thresholds'' derived in other studies, and (particularly) their rapid drop in number fraction at lower redshifts, suggest that massive galaxies are undergoing rapid assembly at $z=3$ followed by a transition to quiescence at $z\lesssim 2$.
\end{itemize}
This study provides an updated description of the evolution of the size-mass relation of galaxies, and the largest samples to date of HST morphologies of massive high redshift galaxies. Additionally, this is the first implementation of the DASH technique for rapid mapping of large areas with HST. We find that the technique is efficient and produces images of nearly the same quality as guided exposures, and encourage further exploitation of this observing mode.

\acknowledgments
This paper is based on observations made with the NASA/ESA Hubble Space Telescope, obtained at the Space Telescope Science Institute, which is operated by the Association of Universities for Research in Astronomy, Inc., under NASA contract NAS 5-26555. These observations are associated with program GO-14114. Support for GO-14114 is gratefully acknowledged. This research made use of Montage. It is funded by the National Science Foundation under Grant Number ACI-1440620, and was previously funded by the National Aeronautics and Space Administration's Earth Science Technology Office, Computation Technologies Project, under Cooperative Agreement Number NCC5-626 between NASA and the California Institute of Technology. J.L.
is supported by an NSF Astronomy and Astrophysics
Postdoctoral Fellowship under award AST-1701487. The Cosmic Dawn Center is funded by the Danish National Research Foundation.

\appendix

\section{COSMOS-DASH Image}

\subsection{Preparation of COSMOS-DASH mosaic}
\label{ap:reduction}

\begin{table}[]
	\centering
	\caption{Data included in the final H$_{160W}$ COSMOS-DASH mosaic.  }
	\begin{tabular}{cccc}
		\hline
		\multicolumn{1}{|c}{\textbf{Program Number}} & \multicolumn{1}{c}{\textbf{Principal Investigator}} & \multicolumn{1}{c}{\textbf{Number of Pointings}} & \multicolumn{1}{c|}{\textbf{Paper}} \\ \hline
		\multicolumn{1}{|c}{\textbf{GO-12167}} & Marijn Franx & 28 & \multicolumn{1}{c|}{\citet{VandeSande2011}} \\
		\multicolumn{1}{|c}{\textbf{GO-12440}} & Sandra Faber & 176 & \multicolumn{1}{c|}{\citet{Grogin2011}} \\
		\multicolumn{1}{|c}{\textbf{GO-12461}} & Adam Reiss & 23 & \multicolumn{1}{c|}{\citet{Oesch2013}} \\
		\multicolumn{1}{|c}{\textbf{GO-12578}} & Natascha F\"orster Schreiber & 112 & \multicolumn{1}{c|}{\citet{Schreiber2013}} \\
		\multicolumn{1}{|c}{\textbf{GO-12990}} & Adam Muzzin & 52 & \multicolumn{1}{c|}{\citet{Marsan2014}} \\
		\multicolumn{1}{|c}{\textbf{GO-13294}} & Alexander Karim & 12 & \multicolumn{1}{c|}{\citet{Gomez-Guijarro2018}} \\
		\multicolumn{1}{|c}{\textbf{GO-13384}} & Dominik Riechers & 4 & \multicolumn{1}{c|}{\citet{Barisic2017}} \\
		\multicolumn{1}{|c}{\textbf{GO-13641}} & Peter Capak & 36 & \multicolumn{1}{c|}{\citet{Barisic2017}} \\
		\multicolumn{1}{|c}{\textbf{GO-13657}} & Jeyhan Kartaltepe & 116 & \multicolumn{1}{c|}{} \\
		\multicolumn{1}{|c}{\textbf{GO-13868}} & Dale Kocevski & 44 & \multicolumn{1}{c|}{} \\
		\multicolumn{1}{|c}{\textbf{GO-14114}} & Pieter van Dokkum & 456 & \multicolumn{1}{c|}{This paper.} \\
		\multicolumn{1}{|c}{\textbf{GO-14895}} & Rychard Bouwens & 20 & \multicolumn{1}{c|}{\citet{Stefanon2017}} \\ \hline
		\textbf{} &  &  &  \\
		\textbf{} &  &  & 
	\end{tabular}
	\label{tab:archival}
\end{table}

Owing to the shifts during the exposures the reduction of DASH data is more complex than that of guided exposures. Here we provide a summary of the reduction procedures; we also refer to \citet{Momcheva2016a} where the reduction of a subset of the COSMOS-DASH data was first described.

Each DASH orbit consists of one guided and seven unguided exposures with offsets of roughly 2$\arcmin$ between each. \citet{Momcheva2016a} describe the overall mosaic strategy and detailed commanding instructions for the DASH visits.  For uniformity of the data reduction, we process both guided and unguided exposures using the same analysis pipeline as outlined below.  

The raw WFC3/IR images were downloaded from the Mikulski Archive for Space Telescopes (MAST\footnote{\url{http://archive.stsci.edu}}) and processed into calibrated exposure ramps (``IMA'' products) with the \texttt{calwf3} pipline after disabling the pipeline cosmic ray identification step (\texttt{CRCORR=False}).  The IMA files provide a measure of the total charge on the detector sampled every 25 seconds over the duration of the exposure (253 or 278 seconds for exposures with \texttt{NSAMP}=12 and 13, respectively).  To reduce the degradation of the image quality of a given exposure due to the telescope drifts, we take \textit{image differences} up the ramp and generate $\mathtt{NSAMP}-2$ essentially independent calibrated exposures with drifts now integrated over 25 seconds rather than the full exposure duration\footnote{Software to split DASH exposures into difference images is provided at https://github.com/gbrammer/wfc3dash}.  The properties of these difference images are essentially identical to normal calibrated WFC3/IR ``FLT'' products, though with slightly different noise characteristics.  Taking image differences increases the effective read noise by a factor of $\sqrt{2}$ and the the read noise of two adjacent image differences will be anti-correlated as the measured (noisy) flux of a given read appears as negative in the first difference image and positive in the second.  In the case of guided exposures with no drifts, the image differences are equivalent to taking the pixel values of the last read minus the first, with read noise from just those two reads.  The differences do not cancel out for sequences with drifts, where the pixel indices of the difference images are effectively shifted when they are combined into the output mosaic.

DASH visits consisting of $\mathtt{NSAMP}-2$ difference image ``exposures'' of a single pointing are then processed in an identical way with guided archival visits consisting of $N$ guided, dithered exposures.  We first compute an internal alignment of the visit exposures using sources detected in the images (both stars and galaxies), which corrects the DASH drifts between samples and small pointing errors typical of the guided sequences. We generate a small mosaic of the visit exposures to detect fainter sources, and align these mosaics to galaxies in the $I_{814W}$ catalogs provided by the COSMOS collaboration \citep{Koekemoer2007TheProcessing}.  Point sources are excluded from the catalog alignment as stars can have significant proper motions between the COSMOS-ACS and DASH epochs.  Since we do not identify cosmic rays in the DASH exposures at the pipeline level as with normal WFC3/IR exposures, we detect and mask the cosmic rays using the standard tools of the \texttt{AstroDrizzle} \citep{Gonzaga2012TheHandbook} package when creating the combined visit/pointing image (turning on cosmic ray identification is useful even for the guided exposures to mask unflagged hot pixels and weaker cosmic rays missed by \textit{calwf3}).  
We note that many sequences of eight pointings were broken up due to South Atlantic Anomaly (SAA) passages. There was typically a large offset of $10\arcsec - 15\arcsec$
between the ``before" and ``after" exposures, due to the spacecraft drift during the SAA passage; nevertheless, no observable degredation of the PSF was found in these sequences. We conclude that in future DASH programs it is not necessary to require avoidance of SAA passages.

Final mosaics consisting of all of the aligned DASH difference images and archival exposures in a given filter were produced with \texttt{astrodrizzle}. We weight the exposures in the final mosaic by their exposure time and drizzle to a pixel scale of $0\farcs 1$ using a square kernel and $\mathtt{pixfrac}=0.8$. The final WFC3/IR mosaic (Fig.~\ref{fig:dash_wht}) spans $9100\times10200$\,pixels, centered at RA$=$10:00:25.4, DEC$=+$2:34:51.2. A list of the archival $H_{160W}$ images which were included in the mosaic are listed in Table \ref{tab:archival}.

\begin{figure*}[htbp]
\centering
\includegraphics[height=3.8in]{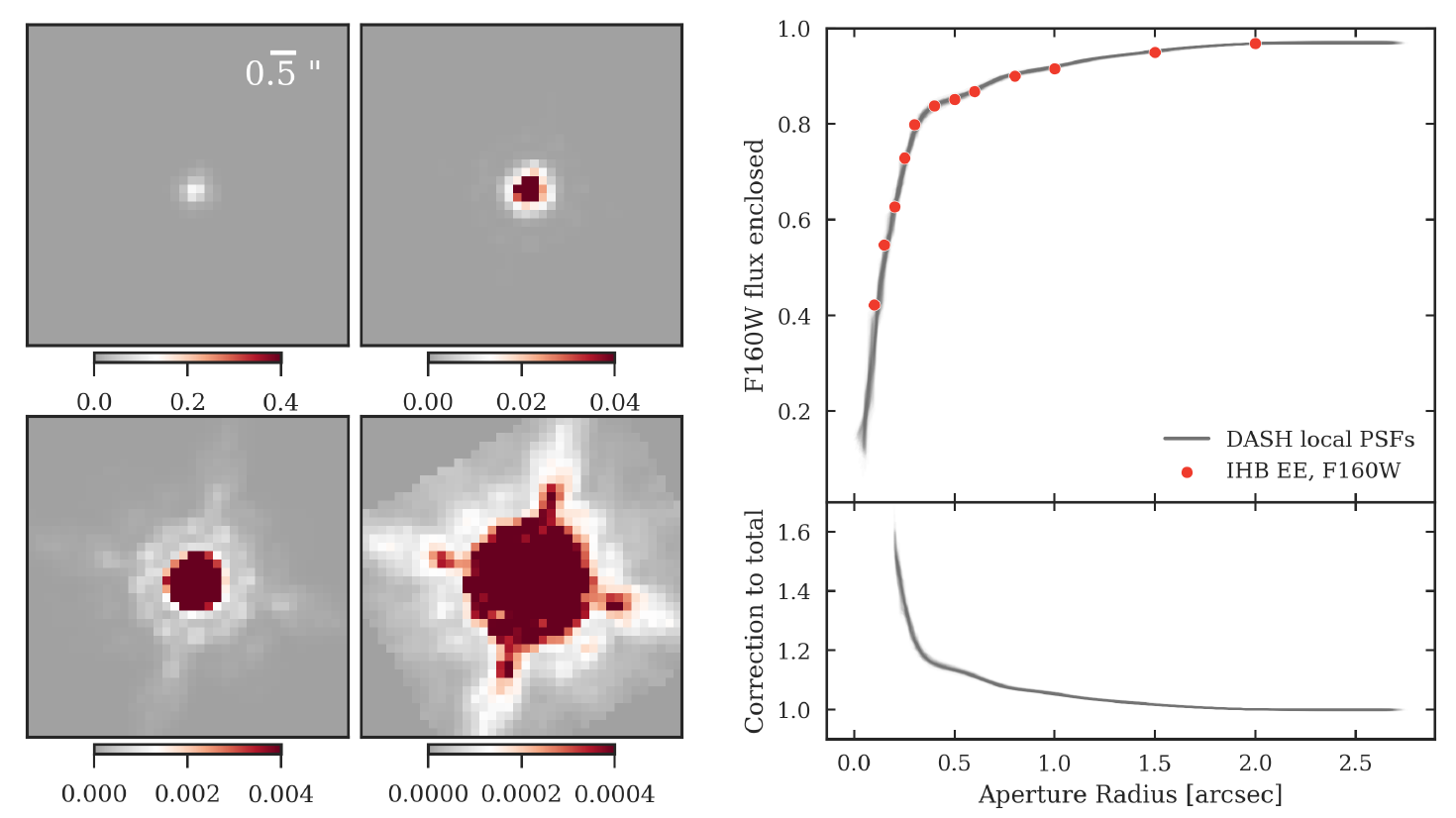}
\caption{The point-spread function (PSF) and growth-curve of COSMOS-DASH. Left: An example local point-spread function (PSFs) in four stretch levels to expose the core, the first Airy ring and the diffraction spikes of the PSF. The construction of the PSFs is described in Section \ref{ap:psf}. Right: $H_{160}$ growth curve. Upper panel shows the fraction of light enclosed as a function of radius to the total light with 2$\arcsec$, $f(r)/f(2\arcsec)$, from all the used $H_{160}$ PSF stamps. The PSFs of the entire field are consistent with each other as shown by the grey lines. The red points show the encircled energy as a function of aperture size, also normalized to 2$\arcsec$, from the WFC3 handbook. The empirical growth curves agree well with the theoretical expectation. The lower panel shows the correction to total flux for point sources across the mosaic with a circularized Kron radius equal to the aperture radius on the x-axis. This is the inverse of the growth curves show in the upper panel ($f(2\arcsec)/f(r)$). The minimum Kron radius is set to 0.3$\farcs$, which requires a maximum correction of 1.6.}
\label{fig:psf}
\end{figure*}

\subsection{Point Spread Function of COSMOS-DASH}
\label{ap:psf}

An example PSF is shown in Fig.\ \ref{fig:psf} at four contrast levels, to demonstrate the different levels of structure: the core of the PSF, the first Airy ring ($\sim$0.5$\%$) and the diffraction spikes ($\sim$0.5$\%$). 
The curves of growth, which show the fraction of light enclosed as a function of aperture size, for the DASH $H_{160}$ PSFs, normalized at 2$\arcsec$, are shown in the left panel of Figure \ref{fig:psf}. The PSFs are in agreement with each other and with the encircled energy as a function of aperture provided in the WFC3 handbook, also normalized to maximum radius of 2$\arcsec$. This demonstrates that the DASH technique does not induce a significant smoothing or other PSF degradation.

\subsection{Background Noise}

\begin{figure*}[ht]
\centering
\includegraphics[width=\textwidth]{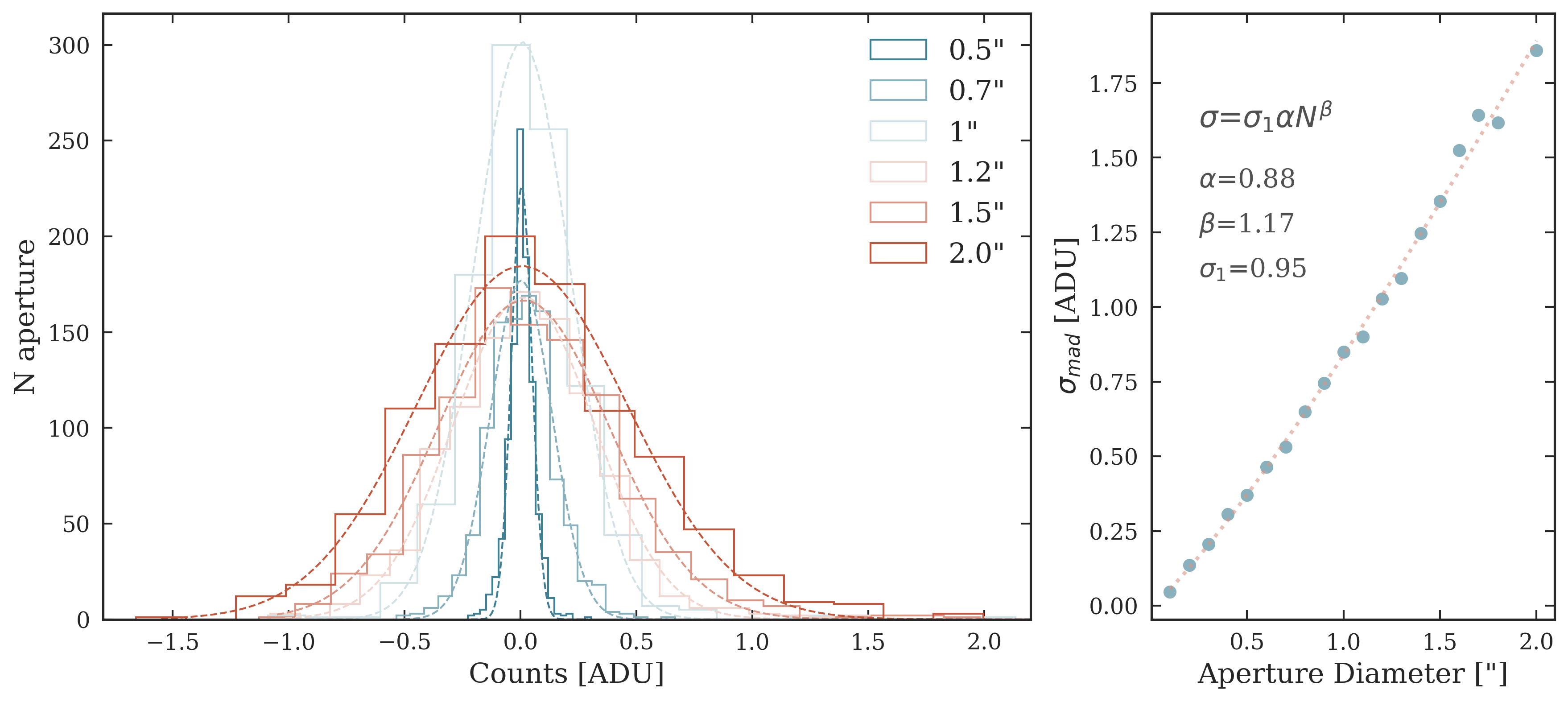}
\caption{Empty aperture photometry on DASH mosaic to determine the background noise level. The left-hand panel shows the histograms of summed counts in different aperture sizes from empty regions throughout the mosaic. Right: Resultant noise-scaling as a function of aperture for the COSMOS-DASH image. The solid red line is a power-law fit to the data. The fit parameters of Eq. \ref{eq:empty_ap} are given in the upper left corner.} 
\label{fig:empty_ap}
\end{figure*}

The depth of the data is determined by the background noise, which is expected to scale with the size of the aperture that is used for photometry.  To determine how the background noise scales with aperture size, we measure the distribution of counts in empty regions of increasing size within the noise-equalized $H_{160}$ image. For each aperture size we measure the flux in $>2000$ apertures placed at random positions across the DASH mosaic. We exclude apertures that overlap with sources in the detection segmentation map. Figure \ref{fig:empty_ap} shows the distribution of of flux counts in empty apertures of 0.5, 0.7, 1., 1.2, 1.5 and 2$\arcsec$ diameters in the DASH $H_{160}$ image. Each histogram can be well-described by a Gaussian, with the width increasing as aperture size increases. The increase in standard deviation with linear aperture size $N=\sqrt{A}$, where A is the area within the aperture, can be described as a power law. A power-law index of 1 would indicate that the noise is uncorrelated, while if the pixels within the aperture were perfectly correlated, the background noise would scale as $N^2$. The right-hand panel of  Figure \ref{fig:empty_ap} shows the measured standard deviation as a function of aperture size in the DASH noise-equalized $H_{160}$ image. We fit a power-law of the form
\begin{equation}
    \sigma = \sigma_1 \alpha N^{\beta},
\label{eq:empty_ap}
\end{equation}

where $\sigma_1$ is the standard deviation of the background pixels fixed to a value of 1.5 here, $\alpha$ is the normalization and $1<\beta<2$ \citep{Whitaker2012}. The fitted parameters are shown in Figure \ref{fig:empty_ap}. The power-law fit is shown by the solid line in the figure.  


\subsection{Zero Point of Mosaic and Point Source Depth}

\begin{figure*}[t]
\centering
\includegraphics[height=3.1in]{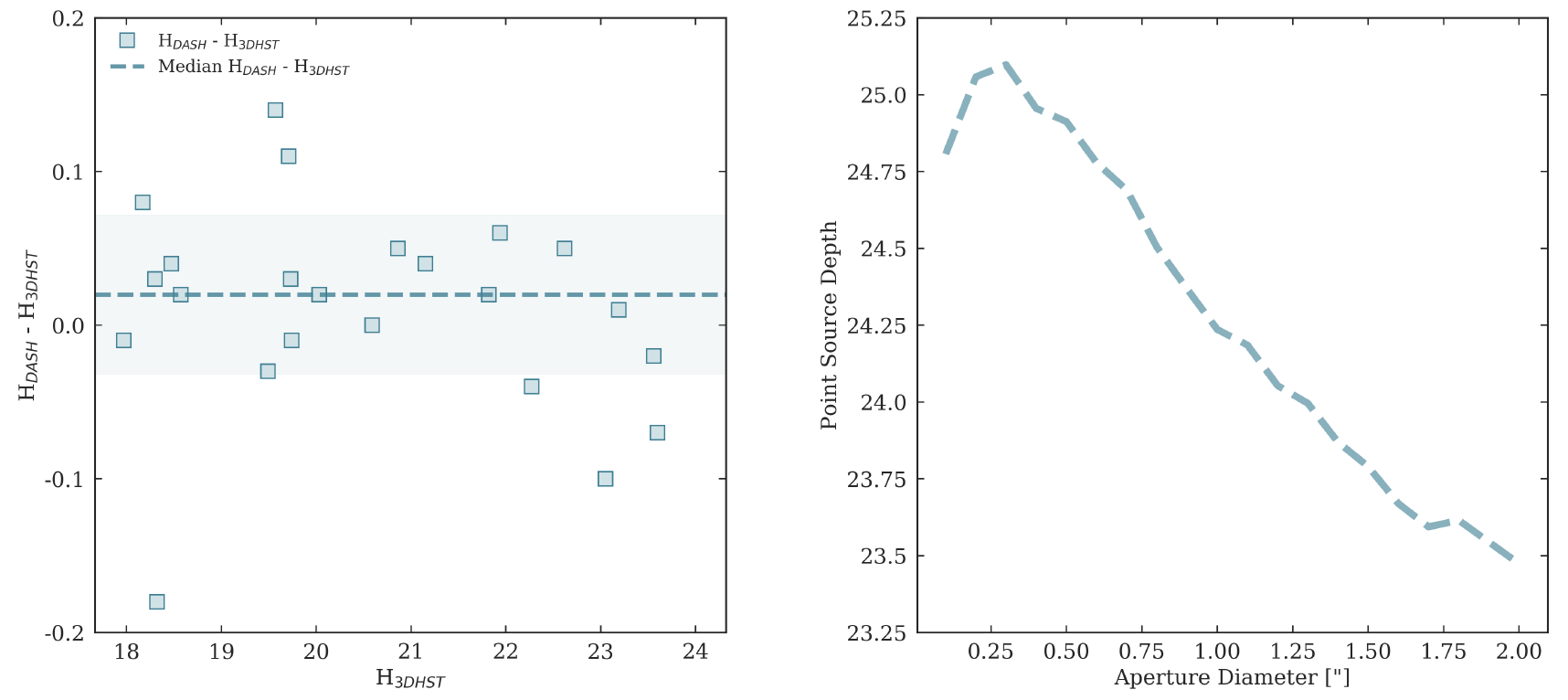}
\caption{Left: Fixed aperture photometry on DASH and CANDELS $H_{160}$ images to determine the zero point of the DASH image. We find the zero-point of the DASH image agrees with the CANDELS image,  and we adopt ZP=25.95. Right: Point source depth of the DASH image, as a function of aperture size. The data reach a depth of 25.1 for the optimal aperture size of $0\farcs 25$, close to the expected value from the ETC of 25.3.}
\label{fig:zp_offset}
\end{figure*}

The DASH technique should not affect the photometric calibration. Nevertheless, we performed an empirical check, making use of the fact that there is a (deliberate) overlap region between the COSMOS-DASH imaging and the CANDELS imaging in the COSMOS field (see \citet{Momcheva2016a} and Fig.\ 1).
We measured the fluxes of stars between $H_{160}=18$ and $H_{160}=24$ in both data sets, adopting the default zeropoint of 25.95 and using identical photometric apertures (of $0\farcs 7$). The results are shown in Fig.\ \ref{fig:zp_offset}; we find that the zeropoint is consistent to within $0.03$ mag.

With the PSF, noise and zeropoint in hand we can calculate the photometric depth of the mosaic.  In the right panel of Fig.\ \ref{fig:zp_offset} we show the $5\sigma$ depth as a function of aperture size. In each aperture, we calculate the noise within the aperture, apply a multiplicative aperture correction based on the flux that falls outside that aperture for a point source, and multiply the resulting number by 5. For very small apertures the S/N is suppressed because of the large aperture corrections that need to be applied, and for very large apertures the S/N is suppressed because of the high noise within the aperture. The optimal aperture is $0\farcs 25$, and the $5\sigma$ depth within that aperture is 25.1. This can be compared to the expected depth for guided exposures of the same exposure time, as determined by the Exposure Time Calculator. The expected depth is
$\approx 25.3$, which means the DASH technique likely imposes a penalty of $\approx 20$\,\% in the flux. This may be because of correlated noise on small scales, due to the shifting of the exposures.

\section{Mass and Size Corrections}

A key goal of our study is to combine our sample with that of \citet{VanderWel2014a}, which was based on the 3D-HST catalogs \citep{Skelton20143D-HSTMasses} in the CANDELS fields \citep{Koekemoer2011}. Here we describe several corrections that we apply to the sizes and masses of the galaxies in order to be on the same ``system" as \citet{VanderWel2014a}. We analyze all galaxies down to $M_*=10^{11}$\,M$_{\odot}$ in the UltraVISTA catalog (i.e., a factor of two below our final mass limit), to ensure that we do not miss any galaxies that have $M_*>2\times 10^{11}$\,M$_{\odot}$ after the corrections.

\subsection{Mass Corrections}
\label{ap:flux_correct}

\begin{figure*}[htbp]
\centering
\includegraphics[width=\textwidth]{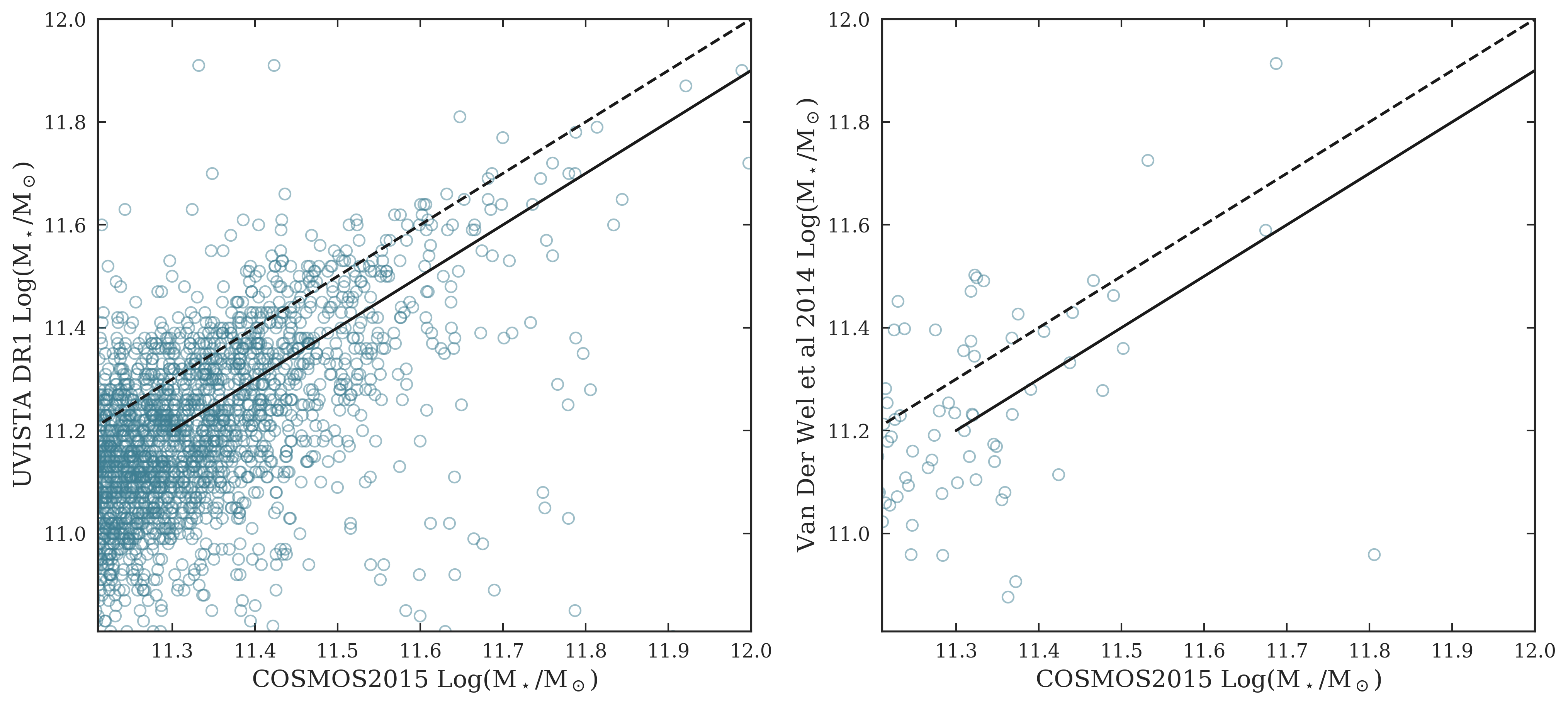}
\caption{Left: Comparison of masses in the \citet{Muzzin2013THESURVEY} UltraVISTA catalog to those in the COSMOS2015 catalog. The 1-1 relation is indicated with a broken line. We find a systematic offset of 0.14 dex (indicated with the solid line). Right panel: comparison of masses in the COSMOS 3D-HST catalog to COSMOS2015. The number of galaxies with $M>2\times 10^{11}$\,M$_{\odot}$ is too small for a secure measurement of the systematic offset.}
\label{fig:mass_corr}
\end{figure*}

We apply two corrections to the stellar masses from the \citet{Muzzin2013THESURVEY} UltraVISTA catalog. The first correction is
a constant offset to bring the masses onto the same system as \citet{VanderWel2014a}. The \citet{Muzzin2013THESURVEY} catalog is based on ground-based data whereas the 3D-HST catalog used by \citet{VanderWel2014a} is selected from HST WFC3 imaging; furthermore, the treatment of the sky background and other steps in the transformation from fluxes to masses can lead to systematic differences. We first ask whether the DR1-based masses of \citet{Muzzin2013THESURVEY} are consistent with the more recent COSMOS2015 catalog from \citet{Laigle2016}. In the left panel of Fig.\ \ref{fig:mass_corr} we show the difference in masses between matched massive galaxies. There is a significant offset; the median offset for galaxies with $M_*({\rm COSMOS2015})>2\times 10^{11}$\,M$_{\odot}$ is $0.14$ dex, such that our masses are smaller than those of COSMOS2015.  In the right panel we show the difference between our masses and those of \citet{VanderWel2014a}. \citet{VanderWel2014a} use masses from the 3D-HST catalogs \citep{Skelton20143D-HSTMasses}. corrected to an infinite aperture using the GALFIT fits (see below). The objects shown in Fig.\ \ref{fig:mass_corr} are objects in the CANDELS/3D-HST COSMOS field, which is inside the UltraVISTA area. There is evidence for a similar offset of $\sim 0.1$ dex at low masses, but at higher masses the COSMOS2015 and \citet{VanderWel2014a} masses appear to be consistent. Unfortunately the number of overlapping objects is too small for a definitive result.

We can derive the mass offset between the \citet{Muzzin2013THESURVEY} UltraVISTA catalog and the \citet{VanderWel2014a} masses in a different way, making use of all five CANDELS fields. The mass function is very steep in this regime, which means that small differences in mass lead to large differences in the number density of galaxies above a particular mass limit. This means we can derive the mass offset by requiring that the number densities of galaxies with $M_*>2\times 10^{11}$\,M$_{\odot}$  in the two surveys are consistent with each other.
The correction to the UltraVISTA masses that we derive this way is $\sim 0.1\,$ dex, in good agreement with the directly measured offset between COSMOS2015 and our UltraVISTA masses. Hence, we applied a $+0.1\,$ dex to all stellar masses of galaxies with $M_{\star}\geq 10^{11} M_{\odot}$ in the \citet{Muzzin2013THESURVEY} catalog. We stress that this correction is uncertain and that it does not necessarily imply that the 3D-HST/\citet{VanderWel2014a} masses are ``more correct" than the \citet{Muzzin2013THESURVEY} masses. We apply the correction to ensure that our data are on the same system as \citet{VanderWel2014a}, so we can meaningfully combine the samples for a joint analysis.

\begin{figure*}[t]
\centering
\includegraphics[width=\textwidth]{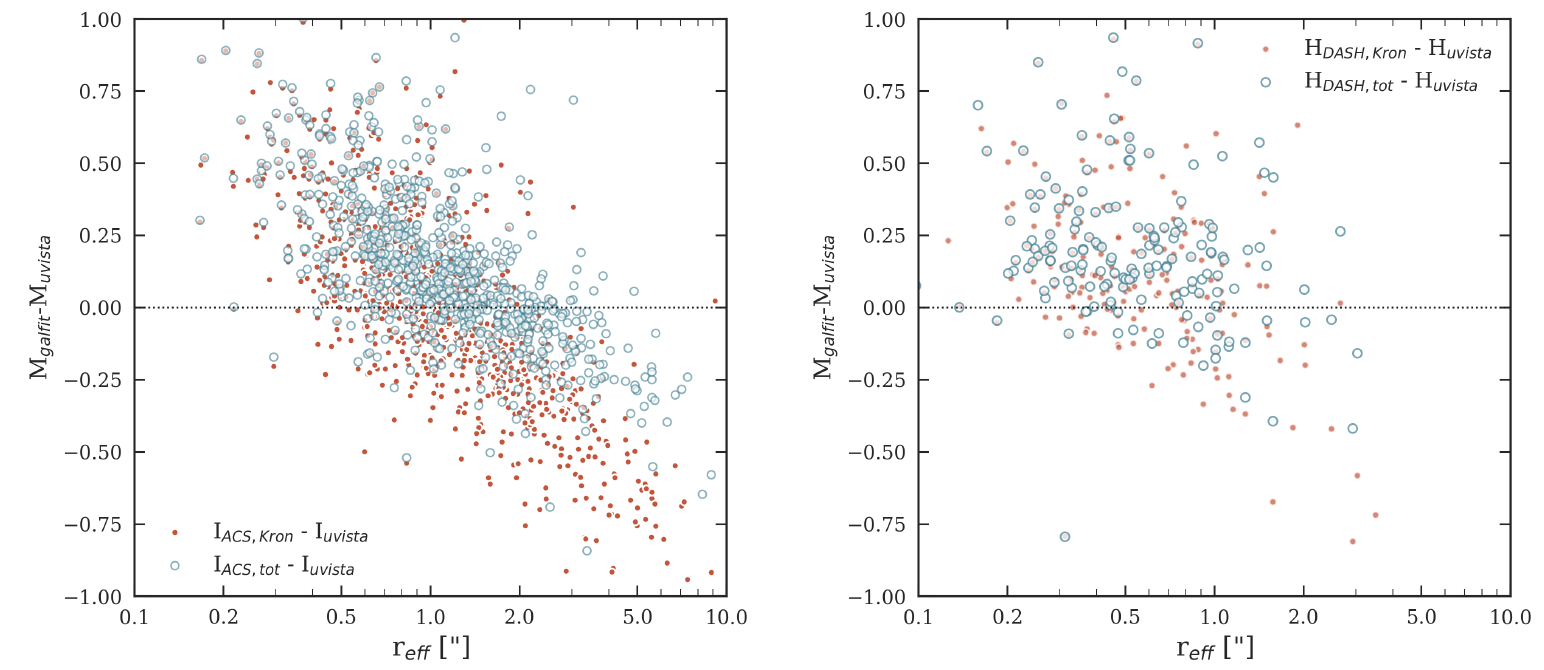}
\caption{Difference between the GALFIT total magnitude and the UltraVISTA magnitude (both AUTO and ``total''), as a function of effective radius. The left panel shows galaxies at $0.1<z<1.5$, where we used the ACS $I_{814}$ imaging, and the right panel shows galaxies at $1.5<z<3.0$ where we used the WFC3 $H_{160}$ images. There are strong correlations, as expected for these large galaxies where a significant fraction of the flux falls outside of the AUTO aperture.}
\label{fig:flux_correct}
\end{figure*}

The second correction is
from the effective aperture in the catalog to $r=\infty$. This correction is based on the GALFIT fits to the galaxies, and ensures that the sizes and masses are self-consistent. This correction is not important at high redshift but significant at low redshift, as these massive galaxies with large S\`ersic indices typically have a large fraction of their flux outside standard photometric apertures. We show this effect in Fig.\ \ref{fig:flux_correct}, for the $I_{814}$ photometry at $0.1<z<1.5$ (left panel) and for the $H_{160}$ photometry at $1.5<z<3.0$ (right panel). The difference between the catalog magnitudes in the corresponding band and the GALFIT total magnitudes is a strong function of the effective radius (in arcseconds), as expected. Red dots are the SExtractor ``AUTO'' magnitudes, and blue dots are ``total'' magnitudes which have been corrected for flux outside of the AUTO aperture. This flux correction is based on point sources, and is insufficient to bring the measurements into agreement with the GALFIT magnitudes.

We correct the masses for this effect in the following way. For each galaxy we convolve the best-fitting GALFIT model with the ground-based $K$-band point spread function (which was used to determine the overall scaling of the SED in the UltraVISTA catalog), and measure the GALFIT model flux inside the AUTO aperture. We then apply the point source-based aperture correction that is listed in the UltraVISTA catalog for the object, and define the mass correction as the ratio between the total GALFIT flux and this aperture-corrected Kron aperture flux. This correction removes the trends seen in Fig.\ \ref{fig:flux_correct}. 

\subsection{Correction for Color Gradients}
\label{ap:wave_correct}

Color gradients and color evolution affect the size measurements of galaxies, as has been found by previous studies \citep{Szomoru2010, VanderWel2014a}. The mass-weighted sizes of galaxies are smaller than the luminosity-weighted sizes, and the sizes in blue bands are generally larger than the sizes in red bands. We follow \citet{VanderWel2014a} and correct the size measurements to a common rest-frame wavelength of 5000\,\AA. This is achieved by measuring the sizes of galaxies in the $I_{814}$ band at $0.1<z<1.5$ and in the $H_{160}$ band at $1.5<z<3.0$, and then applying redshift-dependent corrections. The form of the correction is
\begin{equation}
r_{\rm eff} = r_{\rm eff,F}\left( \frac{1+z}{1+z_p} \right)^\frac{\Delta \log r_{\rm eff}}{\Delta \log \lambda},
\label{eq:m}
\end{equation}
where F denotes either $I_{814}$ (for galaxies at z$\leq$1.5) or $H_{160}$ (for galaxies at z$>$1.5), and $z_p$ is the 'pivot redshift' for these respective filters ($z_p=0.6$ for $I_{814}$ and $z_p=2.2$ for $H_{160}$).

\begin{figure*}[htbp]
\centering
\includegraphics[width=\textwidth]{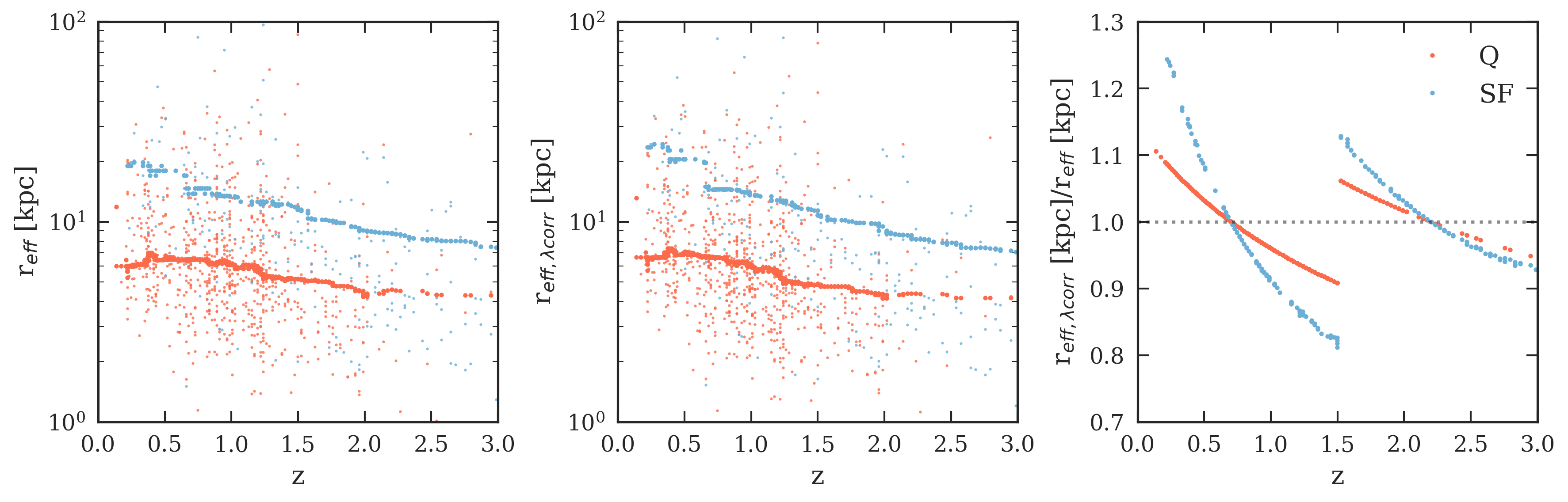}
\caption{Redshift-dependent corrections to the sizes, to correct for color gradients. Left panel: the uncorrected effective radii of star forming (blue) and quiescent (red) galaxies that satisfy our mass selection versus redshift. The lines are running medians. Middle panel: after correction. The evolution is very similar to the uncorrected panel. Importantly, there is no discernable sharp ``feature'' at the redshift where we change from $I_{814}$ sizes to $H_{160}$ sizes. Right panel: the correction for quiescent galaxies and star forming galaxies that is applied. The correction for star forming galaxies is mass-dependent; here it is shown for a mass of $2\times10^{11}$\,M$_{\odot}$.}
\label{fig:wave_corr}
\end{figure*}

Again following \citet{VanderWel2014a},  the color gradient adopted for star forming galaxies is
\begin{equation}
    \frac{\Delta \log r_{\rm eff}}{\Delta \log \lambda} = -0.35 + 0.12z - 0.25\log\left(\frac{M_*}{10^{10}\,{\rm M}_{\odot}}\right)
\end{equation}
and for quiescent galaxies it is simply $\Delta \log$ $r_{\rm eff}/\Delta\log\lambda=-0.25$. These corrections are small, as shown in Fig.\ \ref{fig:wave_corr}.

\subsection{Structural Parameters of All Galaxies Measured by GALFIT}

\begin{figure*}[htbp]
\centering
\includegraphics[width=\textwidth]{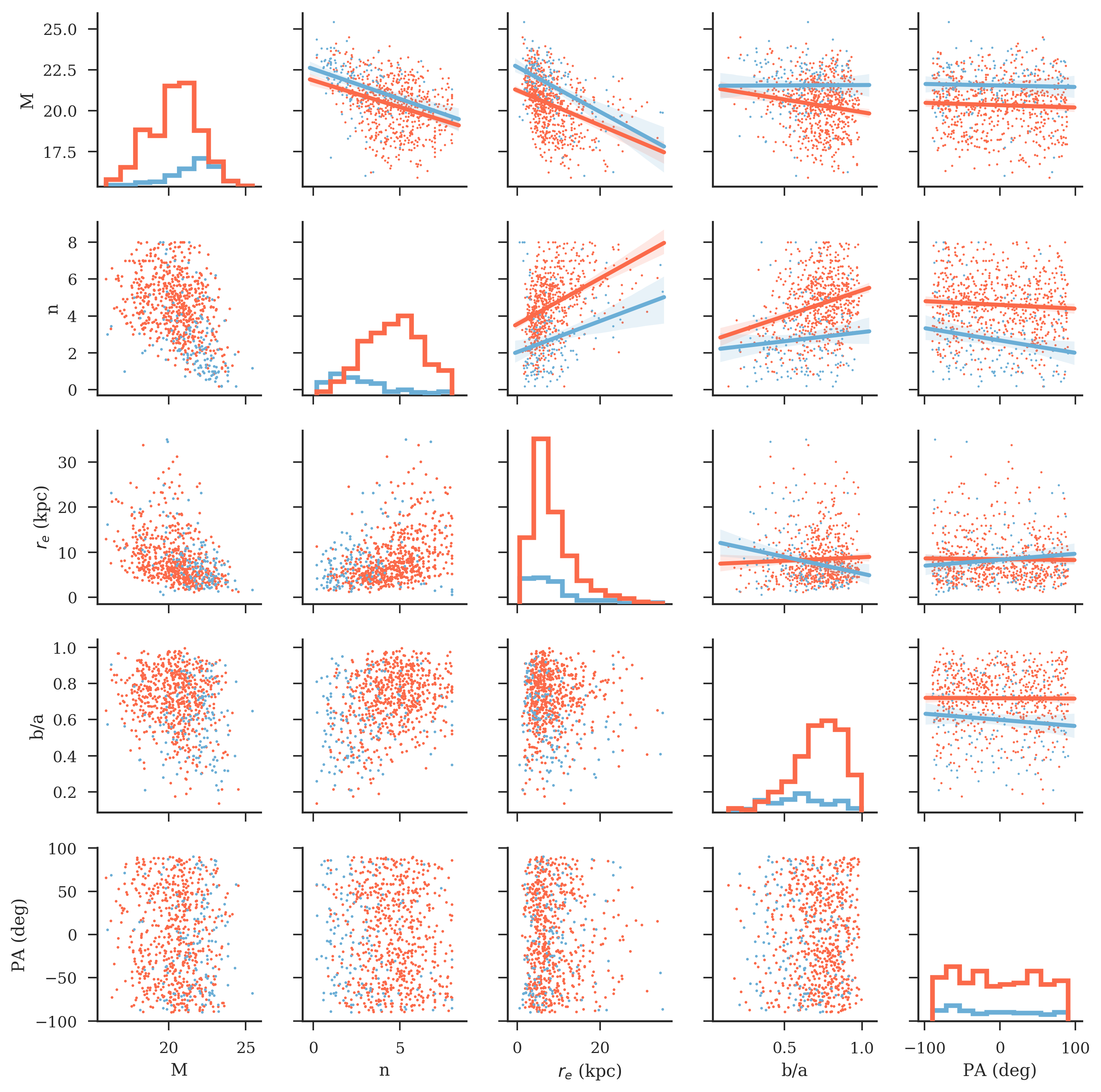}
\caption{The distribution of the GALFIT-derived parameters for the sample of galaxies with $M>2\times 10^{11}$\,M$_{\odot}$: integrated magnitude ($M$) in $H_{160}$ or $I_{814}$ $M$, S\'ersic index ($n$), effective radius ($r_{\rm eff}$), axis ratio ($b/a$) and position angle (PA). The parameters of the star-forming and quiescent galaxies are shown in blue and red respectively. The diagonal panels show histograms of each of the five parameters.} 
\label{fig:galfit_hist}
\end{figure*}

We have measured more information for the sample of massive galaxies than just sizes and fluxes. Although these other parameters are not the main topic of the paper, we provide an overview in Fig.\ \ref{fig:galfit_hist} of the correlations between total magnitude, S\`ersic index, effective radius, axis ratio, and position angle. The position angle is largely a ``control", as no correlations are expected. We see several expected trends, e.g., quiescent galaxies are, on average, rounder than star forming galaxies. It is interesting that the median apparent magnitude of quiescent galaxies is {\em brighter} than that of star forming galaxies, contrary to what one might expect based on their $M/L$ ratios. However, this is simply a reflection of the fact that star forming massive galaxies are only present in large numbers at high redshifts.

\bibliographystyle{apj}
\bibliography{mendeley_v2.bib}

\end{document}